\documentclass[12pt]{article}
\usepackage{latexsym}
\usepackage{amsthm}
\usepackage{amsmath,amssymb,amsfonts,bm,amsbsy,bbm}
\usepackage{epsfig}
\usepackage{graphicx,rotating,lscape}
\usepackage{xcolor}
\usepackage{soul}
\usepackage{hyperref}
\usepackage{verbatim}
\usepackage{natbib}
\usepackage{multirow,color}
\usepackage{geometry}
\usepackage{setspace}
\usepackage{subfigure}
\usepackage{bbm}
\usepackage[flushleft]{threeparttable}
\usepackage{enumitem}
\usepackage{booktabs}
\usepackage{authblk}
\usepackage{titletoc}
\usepackage[ruled,vlined]{algorithm2e}

\newtheorem{assumption}{Assumption} 
\newtheorem{theorem}{Theorem}[section]

\newtheorem{lemma}{Lemma}[section]

\newtheorem{corollary}{Corollary}[section]

\numberwithin{equation}{section}

\usepackage{xr}
\newcommand\ca[1]{{\cal{#1}}}

\newcommand{\indep}{\;\, \rule[0em]{.03em}{.65em} \hspace{-.45em}\rule[-.05em]{.65em}{.03em} \hspace{-.45em}\rule[0em]{.03em}{.65em}\;\,}

\def\wh{\widehat}
\def\wt{\widetilde}
\def\n{\nonumber}   
\def\var{\mbox{var}}
\def\cov{\mbox{cov}}

\def\trans{^{\top}}

\def\sumi{\sum_{i=1}^n}

\def\sumj{{\sum_{j=1}^n}}

\def\ipw{{\mathrm{IPW}}}
\def\orest{{\mathrm{OR}}}
\def\aipw{{\mathrm{AIPW}}}

\def\gsiripw{{\mathrm{GSIR-IPW}}}

\def\gsiraipw{{\mathrm{GSIR-AIPW}}}

\def\gsiraipwcf{{\mathrm{GSIR-AIPW-CF}}}
\def\gsiror{{\mathrm{GSIR-OR}}}

\def\var{\mathrm{var}}

\def\cov{\mathrm{cov}}

\def\0{{\bf 0}}
\def\1{{\bf 1}}

\def\R{{\mathbb{R}}}

\def\b0{{\bf 0}}

\def\mF{\mathcal{F}}

\def\N{\mathbb{N}}

\def\bse{\begin{eqnarray*}}
\def\ese{\end{eqnarray*}}
\def\be{\begin{eqnarray}}
\def\ee{\end{eqnarray}}
\def\bsq{\begin{equation*}}
\def\esq{\end{equation*}}
\def\bq{\begin{equation}}
\def\eq{\end{equation}}

\def\squarebox#1{\hbox to #1{\hfill\vbox to #1{\vfill}}}
\def\boxit#1{\vbox{\hrule\hbox{\vrule\kern6pt\vbox{\kern6pt#1\kern6pt}\kern6pt\vrule}\hrule}}

\definecolor{mygreen}{RGB}{34, 139, 34} 

\allowdisplaybreaks

\title{ Nonlinear Sufficient Dimension Reduction for Causal Inference with Continuous Treatments}
\author[1]{Chak Fung Choi}
\author[2]{Yin Tang}
\author[1]{Bing Li}
\affil[1]{Department of Statistics, The Pennsylvania State University}
\affil[2]{Dr. Bing Zhang Department of Statistics, University of Kentucky}
\date{}

\begin{document}
\maketitle

\begin{abstract}
Causal inference with continuous treatments requires estimating nuisance functions to adjust for the potentially high-dimensional confounder, but nonparametric estimation of these functions is severely hampered by the curse of dimensionality. To address this challenge, we propose a nonlinear sufficient dimension reduction framework to construct low-dimensional sufficient confounders to preserve the unbiasedness of the treatment effect. We estimate the causal effect by combining the generalized Sliced Inverse Regression with the inverse probability weighting, outcome regression, and augmented inverse probability weighting, respectively. 
We further establish the convergence rates of the first two   estimators and derive the asymptotic distributions of the third. Our theoretical results show that, compared to estimators based on the full covariates, dimension reduction yields faster convergence rates and weaker rate conditions without inflating the asymptotic variance. Simulation studies demonstrate that the proposed estimators improve finite-sample performance over their counterparts without dimension reduction, particularly when the original covariate dimension is high. These results demonstrate that sufficient dimension reduction provides a flexible approach to mitigating the curse of dimensionality in causal inference with continuous treatments.
\end{abstract}


\section{Introduction}

Causal inference with continuous treatments arises in many applications involving drug dosage \citep{schomaker2024causal},  political advertising exposure \citep{schindl2026incremental}, environmental exposure \citep{shin2025treatment}, and program intensity \citep{colangelo2025double}. In this setting, the causal effect is naturally described by the dose-response curve, which is the primary target for causal estimation. The dose-response curve can be estimated using various strategies, such as inverse probability weighting, outcome regression, augmented inverse probability weighting, and related doubly robust procedures \citep{kennedy2017non, colangelo2025double}. All of them involving estimation of nuisance functions, such as the generalized propensity score and the conditional mean outcome,  which depend on both the treatment and covariates. Moreover, to ensure all confounding effect are removed, a common practice is to include as many suspicious confounders as possible, resulting in a high-dimensional covariate. This hampers the estimation of the nuisance parameters and consequently also the dose-response curve. The purpose of this paper is to introduce a flexible framework of nonlinear sufficient dimension reduction to overcome the curse of dimensionality caused by the high-dimensional covariate. 

 Specifically, nuisance functions can be estimated using flexible nonparametric and data-adaptive machine-learning methods, including kernel regression \citep{wand1994kernel,fan1996local}, spline smoothing \citep{green1993nonparametric}, random forests \citep{breiman2001random}, neural networks \citep{farrell2021deep,hornik1989multilayer}, as well as ensemble methods such as Super Learner \citep{van2007super}. Estimators of the parameter of interest (dose-response curve) such as double/debiased machine learning \citep{chernozhukov2018double} and targeted maximum likelihood estimation \citep{van2006targeted} can incorporate these flexible nuisance parameter estimators while supporting valid statistical inference under suitable rate conditions, such as rate double-robustness conditions \citep{liu2024assumption,smucler2019unifying}. However, consistency and semiparametric efficiency typically require nuisance function estimators to converge sufficiently fast, a requirement that can be difficult to satisfy with a high-dimensional confounder. This motivates the use of sufficient dimension reduction to facilitate nuisance function estimation.
 
Recently, \cite{luo2017estimating,luo2020matching} applied sufficient dimension reduction (SDR) to reduce the dimension of the confounder to improve accuracy of the estimation of the causal effect; see also \cite{huang2017joint,huang2018joint,ma2019robust}. These methods focused primarily on linear SDR, and can achieve substantial improvement in the accuracy of causal estimation.  A wide variety of linear SDR methods developed in the past two decades or so  and can be applied to the causal estimation setting  \citep{li1991sliced,Cook1991,li-1992,li2007,xia-tong-li-zhu-2002}. See \cite{li2018sufficient,ma2013review} for reviews of this area.
However, in many causal estimation problems, the low-dimensional sufficient predictors are inherently nonlinear, and no reduction is possible when attention is restricted to linear SDR methods. In this paper, we develop a nonlinear SDR method and incorporate it with several most commonly used causal inference methods by replacing the high-dimensional confounder with  our nonlinear low-dimensional sufficient representation of the confounder. 

Other approaches to reducing confounder complexity include covariate selection, balancing-score construction and representation learning. Covariate selection aims to identify a subset of observed covariates that is sufficient for confounding adjustment; see, for example, \cite{brookhart2006variable, de2011covariate, shortreed2017outcome} and \cite{belloni2014inference}. Balancing-score methods summarize the covariate information relevant to treatment assignment through a lower-dimensional score, with examples such as the propensity score for binary treatments \citep{rosenbaum1983central},  the generalized propensity score for continuous treatments \citep{imbens2000role}, and the propensity function framework for general treatment regimes \citep{imai2004causal}. Representation learning uses flexible models such as deep neural networks to learn low-dimensional features that improve covariate balance or counterfactual prediction; see, for example, \cite{johansson2016learning, scholkopf2021toward}.

In this paper we make four main contributions. First, we introduce three types of sufficient confounders—the sufficient confounder, the balancing sufficient confounder, and the prognostic-balancing sufficient confounder—which provide the platform for nonlinear sufficient dimension reduction in causal estimation. Second, we use generalized sliced inverse regression (GSIR) to estimate the balancing sufficient confounder and combine it with inverse probability weighting (IPW) to construct a dimension-reduced causal estimator (GSIR-IPW). Third, we use GSIR to estimate the prognostic-balancing sufficient confounder, then incorporate it into outcome regression (OR) and augmented inverse probability weighting (AIPW) to construct the corresponding dimension-reduced causal estimators (GSIR-OR and GSIR-AIPW). Finally, we establish the convergence rates and asymptotic distributions of the proposed estimators, showing that  they achieve faster rates or attain the same asymptotic variance under weaker rate conditions, compared to their dimension-unreduced counterparts.

The rest of the paper is organized as follows. 
In Section \ref{sec:background} we layout  the background for causal inference under continuous treatments and introduce the three sufficient confounders. 
In Section \ref{sec:GSIR-IPW-OR} introduces the GSIR-IPW and GSIR-OR estimators. In Section \ref{sec:convergence rates GSIR-IPW and GSIR-OR}, we  establish their convergence rates and compare these rates with their dimension-unreduced counterparts.
Sections \ref{sec:GSIR-AIPW} and \ref{sec:gsir-aipw-sec} motivate the GSIR-AIPW estimator from a semiparametric efficiency perspective and develop the corresponding asymptotic theory. 
Section \ref{sec:simulation} reports simulation results assessing the finite-sample performance of the proposed estimators, and Section \ref{sec:application} provides an empirical application. 
Section \ref{sec:discussion} gives some concluding remarks and discussions. Due to space limit, most proofs of the theoretical results are placed in the Supplementary Materials.

\section{Dose-response Curve and Sufficient Confounders}\label{sec:background}
Let $(\Omega, \mF, P)$ be a probability space. We consider an observational study with a continuous treatment and a vector of covariates. The treatment $T$ is a random variable taking values in a measurable space $(\Omega_T, \mF_T)$, where $\Omega_T \subset \R$ is a compact set. The covariate $X$ is a random vector taking values in $(\Omega_X, \mF_X)$, where $\Omega_X \subset \R^p$. Following the potential outcomes framework \citep{rubin1974estimating, imbens2000role}, suppose there exists a collection of potential outcomes $\{Y(t): t \in \Omega_T\}$, where each $Y(t):\Omega \to \R$ is a random variable representing the outcome that would be observed if a subject were assigned treatment level $t$. Since $t$ may not be the treatment level an individual receives, we do not observe $Y(t)$ in practice. Instead, we observe the outcome at the random treatment $T$. Under the commonly used stable unit treatment value assumption (SUTVA), this observed outcome can be written as $Y(T)$. In summary, the observed data consist of $n$ independent and identically distributed (i.i.d.) observations $(Y_i, T_i, X_i)$ for $i = 1,2,\ldots, n$, where $Y_i = Y_i(T_i).$
Our primary estimand is the average dose-response function, defined as 
\bse
\mu(t) = E\{Y(t)\}, \quad \text{for $t\in \Omega_T$},
\ese
which characterizes how the mean potential outcome varies across treatment levels. 

\subsection{Identification of Dose-Response Curve} \label{sec:ident}

We now introduce the assumptions required to identify the dose-response curve from the observed data. In general, for any given $t\in \Omega_T$, $E\{Y(T)|T=t\}$ does not identify $E\{Y(t)\}$, because the potential outcomes $\{Y(t):t \in \Omega_T\}$ is in general not independent of $T$, as the set of covariates may affect both treatment assignment and the potential outcomes. Such covariates form a confounder and must be appropriately adjusted for \citep{vanderweele2013definition}. To ensure that the parameter of interest is identifiable from the observed data, we impose the following assumptions.

\begin{assumption}[Consistency]\label{ass:consistency}
For every $t\in \Omega_T$, $T = t$ implies $Y(T) = Y(t)$.
\end{assumption}
Note that this assumption is already built in the notation $Y(t).$ To see the true meaning of this assumption clearly, let $Y = Y(T)$, then the above assumption should be interpreted as  ``when $T = t$, $Y = Y(t)$." See \cite{hernan2025causal}.
\begin{assumption}[Positivity] \label{ass:positivity}
    $f_{T|X}(t|x) > c$ for some constant $c>0$ for all $(t,x) \in (\Omega_T,  \Omega_X)$, where $f_{T|X}(t|x)$ is the conditional density of $T$ given $X = x$. 
\end{assumption}
\begin{assumption}[Ignorability] \label{ass:ignorability}
    $Y(t) \indep T|X$ for all $t \in \Omega_T$, where $\indep$ denotes independence. 
\end{assumption}
Assumption \ref{ass:consistency} ensures that the 
 observed outcome is exactly the potential outcome corresponding to the subject's own treatment level. Assumption \ref{ass:positivity} states that every subject has some chance of receiving every treatment level regardless of the covariates. It prevents the data from containing regions where some treatment levels can never be assigned under some covariates. Assumption \ref{ass:ignorability}, the commonly used ignorability condition \citep[Chapter 3]{imbens2015causal}, requires that, conditional on the covariates $X$, each potential outcome is independent of the treatment. In other words, adjusting for $X$ removes confounding in the treatment-outcome relationship. 

There are three commonly used approaches that can identify the dose-response curve $E\{Y(t)\}$ under Assumptions \ref{ass:consistency}--\ref{ass:ignorability}. The inverse probability weighting (IPW) identifies it by the relation 
\be 
    E\{Y(t)\} 
    &=& f_T(t) E \left[ \frac{Y(T)}{f_{T|X}(t|X)} \, \bigg| \, T=t \right] \label{eq:ipw-ident-1}\\
    &=&  f_T(t) E \left[ \frac{Y(T)}{f_{T|X}(T|X)} \, \bigg| \, T=t \right].\label{eq:ipw-ident-2}
\ee
Alternatively, the dose-response curve can be identified through the outcome regression (OR) identification formula, 
\begin{equation} \label{eq:or-ident}
    E\{Y(t)\} = E[m(t,X)],
\end{equation}
where $m(t,x) = E\{Y(T) |T = t, X = x\}$. Combining the IPW and OR formulas gives the augmented inverse probability weighting (AIPW) identification formula,
\be
    E\{Y(t)\} &=& E\big[ m(t,X) \big] + f_T(t) E \left[ \frac{Y(T) - m(t,X)}{f_{T|X}(t|X)} \, \bigg| \, T=t \right]  \label{eq:aipw-ident-1}\\
    &=& E\big[ m(t,X) \big] + f_T(t) E \left[ \frac{Y(T) - m(T,X)}{f_{T|X}(T|X)} \, \bigg| \, T=t \right]. \label{eq:aipw-ident-2}
\ee

Conditional on $T = t$, \eqref{eq:ipw-ident-1} and \eqref{eq:ipw-ident-2} are equivalent at the population level, as are \eqref{eq:aipw-ident-1} and \eqref{eq:aipw-ident-2}. However, the corresponding plug-in versions lead to different empirical estimators: \eqref{eq:ipw-ident-1} and \eqref{eq:aipw-ident-1} evaluate the corresponding quantities in the conditional expectation at the target treatment level, while \eqref{eq:ipw-ident-2} and \eqref{eq:aipw-ident-2} evaluate them at each subject's observed treatment level. In practice, the nuisance functions $f_T(t), f_{T|X}(t|x)$, and $m(t,x)$ are replaced by suitable estimators. Unconditional expectations are evaluated using empirical averages, whereas conditional expectations given $T=t$ are estimated using flexible nonparametric or machine learning regression methods.

High-dimensional covariates create substantial challenges for estimating nuisance functions. Many existing estimators for continuous treatments require estimation of the generalized propensity score or related treatment-density quantities, the outcome regression function, or both as functions of the full-covariate vector. For example, \cite{galvao2015uniformly} proposed a semiparametric two-step estimator, while \cite{kennedy2017non} and \cite{colangelo2025double} developed doubly robust estimators. Because these nuisance functions are estimated in the original covariate space, their estimation can become difficult when $X$ is high-dimensional: parametric models may be misspecified, whereas flexible nonparametric estimators may suffer from the curse of dimensionality. These challenges motivate reducing the covariate dimension before estimating the nuisance functions while preserving the information required for causal identification. 

\subsection{Sufficient Confounder} 
\label{sec:sufficient-confounder}

Without further assumptions, adjustment for the full-covariate vector $X$ is needed, because ignorability (Assumption \ref{ass:ignorability}) is assumed to hold only conditional on $X$. Therefore, the nuisance functions, such as the outcome regression function and the generalized propensity score, must be estimated based on the full-covariate vector $X$, which is the
main source of difficulty when $X$ is high-dimensional. We therefore consider the possibility that a low-dimensional structure in $X$ suffices to control for confounding.  That is,  we assume that there exists a function $z: \R^p \to \R^d$ such that 
\be
Y(t) \indep T|z(X) \quad \text{for all} \  t \in \Omega_T. \label{eq:low-dim-ignorability}
\ee
This formulation is more general than variable selection or linear SDR methods, since $z(X)$ may consist of a subset of covariates, or of linear or nonlinear transformation of them. As shown in Theorem \ref{thm:sdr-ident} below, when \eqref{eq:low-dim-ignorability} holds, adjusting for $z(X)$ suffices to identify the dose-response curve from the observed distribution of $\{Y(T),T,z(X)\}$. Thus, although $z(X)$ may have a much lower dimension than $X$, it retains all the information needed to control for confounding. We therefore refer to any transformation satisfying \eqref{eq:low-dim-ignorability} as a \textit{sufficient confounder}.

With a slight abuse of notation, we use
$m(t,x)=E\{Y(T)| T=t,X=x\}$ and
$m(t,z)=E\{Y(T)| T=t,z(X)=z\}$
to denote the outcome regression functions based on the full and
reduced covariates, respectively, with the conditioning covariates
distinguished by the second argument. This notational convention
does not imply that $m(t,x)=m\{t,z(x)\}$, which need not hold for
a general sufficient confounder.

\begin{theorem}\label{thm:sdr-ident}
    Suppose Assumptions \ref{ass:consistency}, \ref{ass:positivity} and \eqref{eq:low-dim-ignorability}  hold for $z(X)$. Then, the IPW identification formulas are
    \be
    E\{Y(t)\} &=& f_T(t) E \left[ \frac{Y(T)}{f_{T|z(X)}\{t|z(X)\}} \, \bigg| \, T=t \right]\label{eq:ipw-ident-sdr-1} \\
    &=& f_T(t) E \left[ \frac{Y(T)}{f_{T|z(X)}\{T|z(X)\}} \, \bigg| \, T=t \right]\label{eq:ipw-ident-sdr-2}.
    \ee
    The OR identification formula is
    \be \label{eq:or-ident-sdr}
    E\{Y(t)\} &=& E[E\{Y(T)|T = t, z(X)\}].
    \ee
    Moreover, the AIPW identification formula is
    \be 
    E\{Y(t)\} &=& E\big[ m\{t,z(X)\} \big] + f_T(t) E \left[ \frac{Y(T) - m\{t,z(X)\}}{f_{T|z(X)}\{t|z(X)\}} \, \bigg| \, T=t \right] \label{eq:aipw-ident-sdr-1} \\
     &=& E\big[ m\{t,z(X)\} \big] + f_T(t) E \left[ \frac{Y(T) - m\{T,z(X)\}}{f_{T|z(X)}\{T|z(X)\}} \, \bigg| \, T=t \right] \label{eq:aipw-ident-sdr-2}.
\ee
\end{theorem}

Theorem \ref{thm:sdr-ident} shows that a sufficient confounder $z(X)$ can replace $X$ in all three identification formulas without changing the target dose-response curve. However, this result does not directly specify how to construct such a sufficient confounder from the observed data or which nuisance function information it should preserve. First, the defining condition \eqref{eq:low-dim-ignorability} involves the potential outcomes $Y(t)$, whereas only $Y(T)$ is observed for each subject. To construct $z(X)$ using SDR, we therefore seek sufficient conditions involving only observed quantities that imply \eqref{eq:low-dim-ignorability}. Second, a sufficient confounder is required only to satisfy  \eqref{eq:low-dim-ignorability}, but the full-covariate generalized propensity score $f_{T|X}(t|x)$  and the outcome regression function $m(t,x)$ are not necessarily preserved. 

We therefore consider two SDR constructions based on the observed data only, motivated by the nuisance functions involved in the respective identification formulas. The first preserves the conditional treatment density and is motivated by IPW; the second additionally preserves the outcome regression and is motivated by OR and AIPW. We introduce these constructions in Sections \ref{sec:balancing} and \ref{sec:prognostic-balancing}, respectively.

\subsection{Balancing Sufficient Confounder}\label{sec:balancing}

We first seek a sufficient confounder adapted to IPW.
Because the IPW identification formula depends on $X$ only through
$f_{T|X}(t|X)$, we seek a low-dimensional transformation
$z(X)$ that preserves this conditional density, that is, $f_{T|X}(t|X) = f_{T|z(X)}\{t|z(X)\}$ for all $t \in \Omega_T$  almost surely.  This is equivalent to the SDR condition
\be
    T \indep X | z(X). \label{eq:T-X-sdr}
\ee
\begin{lemma} \label{lem:sdr1}
Suppose Assumption \ref{ass:ignorability} holds. If $z(X)$ satisfies \eqref{eq:T-X-sdr}, then $z(X)$ satisfies \eqref{eq:low-dim-ignorability}.
\end{lemma}

For binary treatments, a transformation $z(X)$ satisfying
\eqref{eq:T-X-sdr} is called a \emph{balancing score}. 
The propensity score $P(T=1| X)$ is a classical example
\citep{rosenbaum1983central}, since it fully characterizes the
Bernoulli conditional distribution of $T$ given $X$.
For continuous treatments, however, the generalized propensity
score $f_{T|X}(t| x)$ \citep{hirano2004propensity} is indexed
by $t$, and its value at a single treatment level generally
does not determine the full conditional distribution. Following the binary-treatment terminology and in the view of Lemma \ref{lem:sdr1}, we call any $z(X)$ satisfying \eqref{eq:T-X-sdr} a \emph{balancing sufficient confounder}. In contrast to the generalized propensity score, it is a single low-dimensional transformation of $X$ that preserves the entire conditional distribution of $T$ given $X$.

By Theorem \ref{thm:sdr-ident}, 
a balancing sufficient confounder can identify the dose-response curve
through all three identification formulas. Since the generalized propensity score is the only nuisance function involving $X$ in the IPW formula, a balancing sufficient confounder is a natural and parsimonious choice for constructing our IPW estimators. For OR and AIPW estimation, we introduce a second reduction below.

\subsection{Prognostic-Balancing Sufficient Confounder}\label{sec:prognostic-balancing}

Since a balancing sufficient confounder is constructed only to preserve the conditional distribution of $T$ given $X$, it may discard information in $X$ that is relevant to the outcome. In particular, it does not necessarily satisfy 
\be
    m(t,X) = m\{t,z(X)\}, \quad \text{for all $\ t \in \Omega_T$ almost surely. }
    \label{eq:nuis-equiv-1}
\ee
Because the OR and AIPW identification formulas involve the outcome regression function, we additionally seek a reduction that satisfies \eqref{eq:nuis-equiv-1}. However, this equality alone does not guarantee that $z(X)$ is a sufficient confounder. A sufficient condition ensuring both \eqref{eq:T-X-sdr} and \eqref{eq:nuis-equiv-1} is
\be
    \{Y(T),T\} \indep X | z(X). \label{eq:YT-X-sdr}
\ee
 In fact, it is trivial that \eqref{eq:YT-X-sdr} implies \eqref{eq:T-X-sdr}; also, \eqref{eq:YT-X-sdr} implies $Y(T) \indep X |\{T, z(X)\}$, and hence \eqref{eq:nuis-equiv-1}. Thus, a $z(X)$ satisfying \eqref{eq:YT-X-sdr} preserves both nuisance functions, and by Lemma \ref{lem:sdr1}, is also a sufficient confounder. We therefore  use it to construct the OR and AIPW estimators. Moreover, condition \eqref{eq:YT-X-sdr} takes the standard form of an SDR problem, allowing $z(X)$ to be estimated by nonlinear SDR using the observable
joint response $\{Y(T),T\}$.

In fact, condition \eqref{eq:YT-X-sdr} also admits a causal interpretation
through balancing and prognostic properties. We call $z(X)$ a \emph{prognostic sufficient confounder} if
\be \label{eq:prognostic}
    Y(t) \indep X | z(X), \quad \text{for all $t\in\Omega_T$.}
\ee
 This extends the binary-treatment notion of a prognostic score,
defined by $Y(0)\indep X| z(X)$
\citep{hansen2008prognostic}, to all treatment levels. By Lemma \ref{lem:sdr3} in the Supplement, such a $z(X)$ is indeed a sufficient confounder.

\begin{theorem}\label{thm:sdr2} 
Suppose Assumptions \ref{ass:consistency}--\ref{ass:ignorability} hold. Then  the following two statements are equivalent:
\begin{enumerate}
    \item $z(X)$ satisfies both \eqref{eq:T-X-sdr} and \eqref{eq:prognostic};
    \item $z(X)$ satisfies \eqref{eq:YT-X-sdr}.
\end{enumerate}
\end{theorem}

Theorem \ref{thm:sdr2} shows that \eqref{eq:YT-X-sdr} holds if and only if $z(X)$ is both a balancing and a prognostic sufficient confounder. We therefore refer to any $z(X)$ satisfying \eqref{eq:YT-X-sdr} as a \textit{prognostic-balancing sufficient confounder}. That is, the conditional distributions of $T$ and of each $Y(t)$ given $X$ depend on $X$ only through $z(X)$. This result extends that of \cite{huang2018joint}, where $z(X)$ was restricted to be a linear transformation of $X$, namely, $z(X) = B^\top X$ for some $B\in\mathbb{R}^{p\times d}$. Hence, the causal interpretation above applies to any $z(X)$ satisfying \eqref{eq:YT-X-sdr}, including nonlinear transformations of $X$.

\section{GSIR-Based IPW and OR Estimation}  \label{sec:GSIR-IPW-OR}

\subsection{Nonlinear SDR and GSIR} \label{sec:nonlinear-sdr}

Estimating a sufficient confounder $z(X)$ satisfying
\eqref{eq:T-X-sdr} or \eqref{eq:YT-X-sdr} can be formulated as
a nonlinear sufficient dimension reduction (SDR) problem.
For a generic response $U$, nonlinear SDR seeks a low-dimensional
transformation $z:\R^p \to \R^d$, with $d<p$, such that
\be \label{eq:sdr}
U \indep X | z(X).
\ee
The transformed predictor $z(X)$ is called a sufficient predictor for $U$ as it preserves all information in $X$ relevant to $U$. Taking $U=T$ and $U=\{Y(T),T\}$ in \eqref{eq:sdr} gives \eqref{eq:T-X-sdr} and \eqref{eq:YT-X-sdr}, respectively, so the corresponding sufficient predictors are exactly the balancing and prognostic-balancing sufficient confounders.

Existing nonlinear SDR methods include kernel-based approaches, such as GSIR and generalized sliced average variance estimation (GSAVE) \citep{lee2013,li2017nonlinear}, and neural-network-based approaches \citep{liang2022nonlinear,chen2024deep,tang2025belted,xu2025conditional}. Among these choices, we adopt GSIR \citep{lee2013,li2025relative} because its established convergence guarantees \citep{li2017nonlinear,choi2026sharpened} allow us to account for the estimation error in $\wh z$ when analyzing the proposed dose-response estimators. In practice, the estimated sufficient confounder $\wh z(X)$ is obtained by solving a regularized sample eigenvalue problem, following \cite[Section 13.5]{li2018sufficient}. See Appendix \ref{app:nonlinear-sdr} for additional details of nonlinear SDR and GSIR.

\subsection{GSIR-IPW Estimator} \label{sec:IPW}\label{sec:GSIR-IPW}

We first construct kernel-based IPW estimators corresponding to \eqref{eq:ipw-ident-1} and \eqref{eq:ipw-ident-2}, which provide the basis for the GSIR-IPW estimators.
Let the kernel functions be defined as $K_h(t) = h^{-1}K_T(t/h)$ and $K_b(x) = b^{-p}\prod_{k=1}^p K_X(x_k/b)$ for $x = (x_1, \ldots, x_p)\trans$, where $K_T$ and $K_X$ are kernels for $T$ and each component of $X$, respectively, and  $h$ and $b$ are positive bandwidths. Using kernel regression and kernel density estimation, the empirical estimators of the corresponding quantities in \eqref{eq:ipw-ident-1} are
\be\label{eq:ipw-ehat-1}
\wh E \left[ \frac{Y(T)}{f_{T|X}(t|X)} \, \bigg| \, T=t \right]
= \frac{\sumi K_h(t-T_i) Y_i / \wh f_{T|X}(t|X_i) }{\sumi K_h(t-T_i)},
\ee
and
\be\label{eq:f-t-hat}
\wh f_T(t) = \frac{1}{n} \sumi K_h(t-T_i).
\ee
See, for example, \cite{wand1994kernel,tsybakov2009introduction} for more details of the kernel-based estimators.  
To construct the analogous estimators for \eqref{eq:ipw-ident-2}, we replace $\wh f_{T|X}(t|X_i)$ in \eqref{eq:ipw-ehat-1} with $\wh f_{T|X}(T_i|X_i)$ and keep \eqref{eq:f-t-hat}.
Note that, in both IPW estimators, the conditional density can be estimated by the quotient of two unconditional densities as $\wh f_{T|X}(t| x) = \wh f_{T,X}(t,x)/\wh f_X(x)$,
where
\be\label{eq:ipw-nuisance}
\wh f_{T,X}(t,x) = \frac{1}{n}\sumi K_h(t-T_i) K_b(x-X_i), \qquad
\wh f_X(x) = \frac{1}{n}\sumi K_b(x-X_i).
\ee
Combining these estimates yields the two IPW estimators corresponding to \eqref{eq:ipw-ident-1} and \eqref{eq:ipw-ident-2} in the form
\be\label{eq:ipw} 
\begin{split}
&& \wh \mu_{\ipw}^{(t)} (t) 
= \frac{1}{n}\sumi \frac{\wh f_X(X_i)}{\wh f_{T,X}(t,X_i)}Y_i K_h(t-T_i),\ \\
&& \wh \mu_{\ipw}^{(T)} (t) 
= \frac{1}{n}\sumi \frac{\wh f_X(X_i)}{\wh f_{T,X}(T_i,X_i)}Y_i K_h(t-T_i).
\end{split}
\ee
Throughout this section, for simplicity, we assume that the same kernel and bandwidth are used for $T$ in both \eqref{eq:ipw-nuisance} and \eqref{eq:ipw}, where the kernel for $T$ is used for local regression in the former and for density estimator in the latter. In practice, these kernels and bandwidths may be different, as illustrated by the AIPW estimator introduced in Section \ref{sec:GSIR-AIPW}. 

When a low-dimensional balancing sufficient confounder $z(X)$ satisfying \eqref{eq:T-X-sdr} exists, we construct two GSIR-IPW estimators based on \eqref{eq:ipw-ident-sdr-1} and \eqref{eq:ipw-ident-sdr-2}. The first step is to apply GSIR with response $T$ and predictor $X$ to obtain an estimate $\wh z(x)=\{\wh z_1(x),\ldots,\wh z_d(x)\}^\top$ of $z(x)$, and define $\wh Z_i=\wh z(X_i)$ for $i=1,\ldots,n$. The second step is to replace $X_i$ with $\wh Z_i$ in \eqref{eq:ipw}. Alternatively, we may apply GSIR with response $\{Y(T),T\}$ and predictor $X$ to target \eqref{eq:YT-X-sdr}; the subsequent IPW construction is unchanged.

Specifically, the resulting GSIR-IPW estimators are given by
\be\label{eq:gsir-ipw-1} 
\wh \mu_{\gsiripw}^{(t)} (t) 
= \frac{1}{n}\sumi \frac{\wh f_Z(\wh Z_i)}{\wh f_{T,Z}(t,\wh Z_i)}Y_i K_h(t-T_i),
\ee
and
\be\label{eq:gsir-ipw-2} 
\wh \mu_{\gsiripw}^{(T)} (t)
= \frac{1}{n}\sumi \frac{\wh f_Z(\wh Z_i)}{\wh f_{T,Z}(T_i,\wh Z_i)}Y_i K_h(t-T_i),
\ee
where the two density functions are estimated by
\bse\label{eq:kernel-density}
\wh f_{T,Z}(t,z) = \frac{1}{n}\sumj K_h(t-T_j) K_b(z-\wh Z_j), \qquad
\wh f_Z(z) = \frac{1}{n}\sumj K_b(z-\wh Z_j).
\ese
Note that, by an abuse of notation, we use the same symbol $K_b$ to denote $K_b(z) = b^{-d}\prod_{k=1}^d K_Z(z_k/b)$ for $z=(z_1,\dots,z_d)\trans$, where $K_Z$ is the kernel for each component of $Z$.

\subsection{GSIR-OR Estimator}\label{sec:OR}  \label{sec:GSIR-OR}
In parallel to Section \ref{sec:IPW}, we first construct the OR estimator, which provides the basis for the GSIR-OR estimator. Replacing the expectation in \eqref{eq:or-ident} by a sample average and $m(t,x)$ by its Nadaraya-Watson estimator $\wh m(t,x)$ \citep{wand1994kernel,tsybakov2009introduction} gives
\be\label{eq:or} 
\qquad \wh \mu_{\orest}(t)
= \frac{1}{n}\sum_{i=1}^n \wh m(t,X_i), 
\quad \text{where} \quad
\wh m(t,x) = \frac{\sumj Y_j K_h(t-T_j) K_b(x-X_j)}{\sumj K_h(t-T_j) K_b(x-X_j)}.
\ee

As in Section \ref{sec:GSIR-IPW}, we use the identity \eqref{eq:or-ident-sdr} to construct the GSIR-OR estimator in two steps. In the first step, we implement GSIR with response $\{Y(T),T\}$ and predictor $X$ to estimate a prognostic-balancing sufficient confounder satisfying \eqref{eq:YT-X-sdr}. This condition is stronger than the sufficient condition \eqref{eq:T-X-sdr} for GSIR-IPW estimation and additionally ensures \eqref{eq:nuis-equiv-1}. Therefore, the outcome regression can be estimated  using the reduced sufficient confounder $z(X)$. Let $\wh z$ denote the resulting estimate and define $\wh Z_i=\wh z(X_i)$ for $i=1,\ldots,n$. In the second step, we estimate $m(t,z) $ using kernel regression.  Replacing $X_i$ and $X_j$ in \eqref{eq:or} with $\wh Z_i$ and $\wh Z_j$, respectively, gives the GSIR-OR estimator
\be\label{eq:gsir-or}
\qquad\qquad  \wh\mu_{\gsiror}(t) = \frac{1}{n}\sum_{i=1}^n \wh m(t,\wh Z_i),
\quad \text{where} \quad
\wh m(t,z) = \frac{\sumj Y_j K_h(t-T_j) K_b(z-\wh Z_j)}{\sumj K_h(t-T_j) K_b(z-\wh Z_j)}.
\ee

\section{Convergence Rates of GSIR-IPW and GSIR-OR}\label{sec:convergence rates GSIR-IPW and GSIR-OR}

\subsection{Convergence Rate of GSIR-IPW} \label{sec:gsir-ipw-rate}

Let $r_n \to 0$ be a deterministic sequence characterizing the
uniform convergence rate of the GSIR estimator $\wh z$, such that
\be\label{eq:gsir-rate-rn}
\sup_{x \in \Omega_X}\|\wh z(x)-z(x)\|_2 = O_p(r_n).
\ee
The following two theorems establish the asymptotic properties
of the GSIR-IPW estimators defined in \eqref{eq:gsir-ipw-1}
and \eqref{eq:gsir-ipw-2}, accounting for the estimation error
in $\wh z$ through $r_n$. See Appendix~\ref{app:gsir-rate} for further discussion of $r_n$
and the choice of tuning parameters yielding the optimal convergence rate.

\begin{theorem}\label{thm:gsir-ipw-rate-1}
Suppose Assumptions \ref{ass:fZ-bound}--\ref{ass:rate} hold and the GSIR estimator $\wh z$ satisfies \eqref{eq:gsir-rate-rn}. Let 
\be\label{eq:gsir-ipw-rate}
s_n &=& n^{-1/2} h^{-1/2} + r_nb^{-1} + r_n^2 b^{-d-2}h^{-1} + b^s + h^q + n^{-1} b^{-d}h^{-1}\log n \n\\
&& + r_n n^{-1}h^{-1}b^{-(d+1)} \{\log(1/b) + \log(1/h)\} \n\\
&&+ r_n n^{-1/2} h^{-1/2} b^{-(d+2)/2}  \sqrt{\log(1/b) + \log(1/h)}.
\ee
Suppose $h$ and $b$ are selected such that $s_n \to 0$ as $n \to \infty$. Then, for an arbitrary $t$, the estimator $\wh \mu_{\gsiripw}^{(t)} (t)$ given by \eqref{eq:gsir-ipw-1} is a consistent estimator of $\mu(t)$, with the convergence rate
\bse
 \wh \mu_{\gsiripw}^{(t)} (t) - \mu(t) = O_p(s_n).
\ese

\end{theorem}
\begin{theorem}\label{thm:gsir-ipw-rate-2}
 Suppose that all assumptions in Theorem \ref{thm:gsir-ipw-rate-1} hold, and in addition $|K_T|*|K_T|$ is square integrable (i.e., $\int \{(|K_T|*|K_T|)(u)\}^2 du < \infty$, where $*$ refers to convolution). Then, for an arbitrary $t$, $\wh \mu_{\gsiripw}^{(T)} (t)$ given by \eqref{eq:gsir-ipw-2} is a consistent estimator for $\mu(t)$, with the convergence rate
\bse
 \wh \mu_{\gsiripw}^{(T)} (t) - \mu(t) = O_p(s_n),
\ese
provided that $s_n \to 0$ as $n\to \infty$, where $s_n$ is given by \eqref{eq:gsir-ipw-rate}.
\end{theorem}

Hence, we derive their optimal rate based on the representation $s_n$ in \eqref{eq:gsir-ipw-rate},
when $h \asymp n^{-\gamma}$, $b \asymp n^{-\eta}$ and $r_n \asymp n^{-\delta}$. Note that $\delta$ depends on the GSIR operators and the choice of tuning parameters. We assume that these parameters are chosen optimally and hence treat $\delta$ as fixed here.

\begin{theorem}\label{thm:optimal-rate-ipw-gsir}
Suppose that the tuning parameters, the bandwidths of two kernel functions, have the form $h \asymp n^{-\gamma}$ and $b \asymp n^{-\eta}$ for some $\gamma,\eta>0$. Also, assume the GSIR rate is $r_n \asymp n^{-\delta}$, where $0 < \delta \le 1$ is fixed. Then, the optimal convergence rate (up to a logarithmic factor) of  $s_n$ in \eqref{eq:gsir-ipw-rate} is $n^\theta$, where
\be\label{eq:theta-optimal-ipw-gsir}
\theta = \max\left\{-\frac{q}{1+2q}, -\frac{s \delta}{s+1}, - \frac{2s \delta q}{q(d+s+2) + s}, -\frac{sq}{sq+dq+s}\right\}.
\ee
\end{theorem}

\subsection{Convergence Rate of GSIR-OR} \label{sec:gsir-or-rate}
The following theorem provides the asymptotic analysis of the GSIR-OR estimator in \eqref{eq:gsir-or}. 

\begin{theorem}\label{thm:gsir-or-rate}
Under all assumptions in Theorem \ref{thm:gsir-ipw-rate-1}, for an arbitrary $t$, $\wh \mu_{\gsiror} (t)$ defined in \eqref{eq:gsir-or} is a consistent estimator for $\mu(t)$, with the convergence rate
\bse
\wh \mu_{\gsiror} (t) - \mu(t) = O_p(s_n),
\ese
 provided that $s_n \to 0$ as $n \to \infty$, where $s_n$ is given by \eqref{eq:gsir-ipw-rate}.
\end{theorem}

Since the above convergence rate is the same as that of the GSIR-IPW  estimators in Theorems \ref{thm:gsir-ipw-rate-1} and \ref{thm:gsir-ipw-rate-2}, the optimal convergence rate in Theorem \ref{thm:optimal-rate-ipw-gsir} also applies to the GSIR-OR estimator under the same conditions, and is hence omitted.

\subsection{Comparison of Optimal Rates with Full-Covariate Estimators}

The analogous asymptotic results for the IPW and OR estimators
without SDR are provided in the Supplementary Material; see details in Appendix \ref{app:ipw-or-rate}. 
The convergence rate analogous to $s_n$ in \eqref{eq:gsir-ipw-rate} is denoted by $s_n'$ in \eqref{eq:ipw-rate}, and its corresponding optimal rate is denoted by $n^{\theta'}$, where $\theta'$ is given by \eqref{eq:theta-optimal-ipw}.
Here, we compare their optimal convergence rates with those
of the GSIR-based estimators.

To compare the derived rate bounds with and without GSIR, we fix $q=s=2$, corresponding to the order of Gaussian kernels, and evaluate $\theta$ and $\theta'$ in \eqref{eq:theta-optimal-ipw-gsir} and \eqref{eq:theta-optimal-ipw} for $p=5,10,15,20,25,30$, $d=1,2$, and $\delta=1/3,7/24,1/4$. We include $\delta=1/3$ as a limiting benchmark corresponding to the convergence exponent that GSIR can approach \citep{choi2026sharpened}, $\delta=1/4$ as the exponent established by \citep{li2017nonlinear} under their most favorable smoothness conditions, and $\delta=7/24$ as an intermediate value. We consider $d=1,2$ as illustrative reduced dimensions, motivated by the commonly-used regression examples in \citep[Section S.5.2]{tang2025belted}. The resulting comparisons of $\theta$ and $\theta'$ are reported in Table \ref{tab:rate-comparison}.

Several observations can be made from Table \ref{tab:rate-comparison}. First, when the original covariate dimension is small ($p=5$), GSIR does not improve the convergence rate for any value of $\delta$ considered. This is expected because the gain from dimension reduction is not sufficient to compensate for the additional estimation error introduced by GSIR. In contrast, once $p \ge 10$, GSIR achieves a faster convergence rate in all settings considered, except for $\delta = 1/4, d= 2$ and $p = 10$. As expected, the benefit of GSIR becomes more apparent as $p$ increases.

Table \ref{tab:rate-comparison} also shows that this qualitative conclusion is robust to changes in the GSIR convergence rate. Although a smaller value of $\delta$ (slower GSIR convergence rate) slightly reduces the improvement in the convergence rate of the dose-response estimator, the threshold at which GSIR becomes beneficial remains unchanged in most settings considered, shifting from $p=10$ to $p=15$ only when $\delta=1/4$ and $d=2$.  Furthermore, increasing the reduced dimension from $d=1$ to $d=2$ leads to a smaller improvement, since the reduction in dimensionality is less substantial. 

Finally, it should be emphasized that the convergence rates reported in Table \ref{tab:rate-comparison} are theoretical upper bounds. In finite samples, the actual performance of the GSIR-based estimator also depends on factors such as the constants hidden in the asymptotic rates, the estimation accuracy of the sufficient confounders, and the choice of tuning parameters.

\section{GSIR-Based AIPW Estimation} \label{sec:GSIR-AIPW} 

\subsection{GSIR-AIPW Estimator}
Although the pointwise dose-response function $\mu(t)$ has no efficient influence function (EIF) in a fully nonparametric model, the weighted integral $\int w(u)\mu(u)\,du$ can admit one under suitable regularity conditions. While  previous work considers weighting by the marginal treatment density $f_T$ \citep{kennedy2017non}, we derive the EIF for a general fixed weight function $w$ in Appendix \ref{app:aipw-motivation}. The resulting augmentation structure motivates two AIPW estimators that combine outcome regression with inverse probability weighting. 
Under suitable regularity conditions, their population
counterparts recover \eqref{eq:aipw-ident-1} and
\eqref{eq:aipw-ident-2}, respectively, as $h\to 0$:
\be
\wh\mu_{\aipw}^{(t)}(t)
&=& \frac{1}{n}\sumi
\left[
\wh m(t,X_i)
+ \frac{K_h(T_i-t)}{\wh f_{T|X}(t|X_i)}
\{Y_i-\wh m(t,X_i)\}
\right],
\label{eq:aipw-1}\\
\wh\mu_{\aipw}^{(T)}(t)
&=& \frac{1}{n}\sumi
\left[
\wh m(t,X_i)
+ \frac{K_h(T_i-t)}{\wh f_{T|X}(T_i|X_i)}
\{Y_i-\wh m(T_i,X_i)\}
\right].
\label{eq:aipw-2}
\ee
Here, and $\wh m$ and $\wh f_{T|X}$ may be obtained using nonparametric or machine learning methods.
Estimator \eqref{eq:aipw-1} was previously considered by
\cite{colangelo2025double} using cross-fitting.
In fact, both estimators are doubly robust up to the smoothing bias caused by $K_h$, as stated in Proposition \ref{prop:doubly-robust} in Appendix \ref{app:doubly-robust}.

As in Sections \ref{sec:GSIR-IPW} and \ref{sec:GSIR-OR}, we construct the GSIR-AIPW estimators in two steps. In the first step, we apply GSIR with response $\{Y(T),T\}$ and predictor $X$ to estimate a prognostic-balancing sufficient confounder satisfying \eqref{eq:YT-X-sdr}. Denote the estimator by $\wh z$ and define $\wh Z_i=\wh z(X_i)$ for $i=1,\dots,n$. Since \eqref{eq:YT-X-sdr} implies \eqref{eq:T-X-sdr} and \eqref{eq:nuis-equiv-1}, $z(X)$ preserves both nuisance functions required for AIPW estimation. In the second step, we replace $X_i$ with $\wh Z_i$ in \eqref{eq:aipw-1} and \eqref{eq:aipw-2} to obtain the GSIR-AIPW estimators
\be
\wh \mu_{\gsiraipw}^{(t)} (t) &=& \frac{1}{n} \sumi \left[\wh m(t,\wh{Z}_i) + \frac{K_h(T_i-t)}{\wh f_{T|Z}(t|\wh{Z}_i)}\{Y_i - \wh m(t, \wh{Z}_i)\} \right]\label{eq:gsir-aipw-1},\\
\wh \mu_{\gsiraipw}^{(T)} (t) &=& \frac{1}{n} \sumi \left[\wh m(t,\wh{Z}_i) + \frac{K_h(T_i-t)}{\wh f_{T|Z}(T_i|\wh{Z}_i)}\{Y_i - \wh m(T_i, \wh{Z}_i)\} \right] \label{eq:gsir-aipw-2}.
\ee

\subsection{Sample Splitting and Cross-Fitting} \label{sec:cross-fitting}

To obtain valid inference with flexible nonparametric nuisance function estimators, we use the following $L$-fold cross-fitting procedure to separate nuisance estimation from evaluation of the estimating equation.
\begin{enumerate}
\item[Step 1:] For some fixed integer $L \ge 2$, randomly partition the observation indices into $L$ distinct groups, each with index set $I_l$, for $l = 1, \ldots, L$, such that the sample size of each group is $n_l$. For simplicity, we assume that $n_l = n/L$ and that $n$ is divisible by $L$.
\item[Step 2:]
For each $l$, we randomly split the observations not in $I_l$ into two approximately equal-sized subsets, $J_{l,1}$ and $J_{l,2}$, with sample sizes $n_{l,1}$ and $n_{l,2}$, respectively. We use the observations in $J_{l,1}$ to estimate  a prognostic-balancing sufficient confounder $z(X)$ satisfying \eqref{eq:YT-X-sdr}, and denote the estimator by $\wh z_l(\cdot)$. Next, we apply $\wh z_l(\cdot)$ to the observation in $J_{l,2}$ and $I_l$ to obtain $\wh Z_{li} = \wh z_l(X_{i}), i \in J_{l,2} \cup I_l$. We then use $\{(Y_i, T_i, \wh Z_{li}): i \in J_{l,2}\}$ to estimate $m(t,z)$ and $f_{T|Z}(t| z)$, denoted by $\wh m_l(t,z)$ and $\wh f_{l,T| Z}(t| z)$, respectively. 

\item[Step 3:]  Define the cross-fitted GSIR-AIPW estimators of the dose-response function corresponding to \eqref{eq:gsir-aipw-1} and \eqref{eq:gsir-aipw-2} as follows:
\end{enumerate}
\be
\wh \mu_{\gsiraipwcf}^{(t)}(t) &=& \frac{1}{n}  \sum_{l=1}^L  \sum_{i \in I_l} \left[ \wh m_l(t, \wh Z_{li}) + \frac{K_h(t-T_i)}{\wh f_{l, T| Z}(t|\wh Z_{li})}\{Y_i - \wh m_l(t, \wh Z_{li})\} \right] \label{eq:gsir-aipw-estimator-1},\\
\wh \mu_{\gsiraipwcf}^{(T)}(t) &=& \frac{1}{n}\sum_{l=1}^L  \sum_{i \in I_l} \left[\wh m_l(t, \wh Z_{li}) + \frac{K_h(t-T_i)}{\wh f_{l, T|  Z}(T_i|\wh Z_{li})}\{Y_i - \wh m_l(T_i, \wh Z_{li})\} \right] \label{eq:gsir-aipw-estimator-2}.
\ee

In fact, as pointed out by \cite{kennedy2024semiparametric,kennedy2017non}, the nuisance function estimators $\wh m_l(t,z)$ and $\wh f_{l,T|Z}(t|z)$ in \eqref{eq:gsir-aipw-estimator-1} and \eqref{eq:gsir-aipw-estimator-2} can be constructed using flexible nonparametric or machine learning methods. As we will see in Section \ref{sec:gsir-aipw-asymptotic} below, some rate conditions will be required in order to guarantee the asymptotic normality of the estimators, and some rate conditions based on kernel estimators will be discussed in Section \ref{sec:discussion-gsir-aipw}.
\section{Asymptotic Analysis for GSIR-AIPW} \label{sec:gsir-aipw-sec}
\subsection{Asymptotic Distribution of GSIR-AIPW}\label{sec:gsir-aipw-asymptotic}

We now present the asymptotic analysis for the two GSIR-AIPW estimators \eqref{eq:gsir-aipw-estimator-1} and \eqref{eq:gsir-aipw-estimator-2} under cross-fitting. 
For notational convenience, denote
\bse
\wh q_l(t,z) = \frac{1}{\wh f_{l,T|Z}(t|z)} =\frac{\wh f_{l, Z}(z)}{\wh f_{ l, T, Z}(t, z)},\quad
\text{and} \quad 
q(t,z) = \frac{1}{f_{T| Z}(t | z)} = \frac{f_Z(z)}{f_{T, Z}(t, z)}.
\ese

We first give the theorem for the asymptotic expansion of $\wh \mu_{\gsiraipwcf}^{(t)}(t)$ given by \eqref{eq:gsir-aipw-estimator-1}, followed by an asymptotic normality result.

\begin{theorem} \label{thm:gsir-aipw-1}
Suppose that Assumptions \ref{ass:fZ-bound}, \ref{ass:KT-bound}, \ref{ass:Y-bound} and \ref{ass:diff-bound} hold, $h \to 0$ and $nh \to \infty$.  Further suppose that the convergence rates of the estimators $\wh m_l(t, \wh z_l)$ and $\wh q_l(t,\wh z_l)$, defined by
\be\label{eq:m-q-vn}
\sup_{t, z, \wh z} |\wh m_l(t, \wh z_l) - m(t, z)| = O_p(u_n), 
\quad \text{and} \quad
\sup_{t, \wh z, z} |\wh q_l(t,\wh z_l) - q(t,z)| = O_p(v_n),
\ee
for $l=1,\dots,L$, satisfy $u_n \to 0$, $v_n \to 0$, $n^{1/2} h^{1/2} u_n v_n \to 0$, $n^{1/2} h^{q+1/2} u_n \to 0$, and $n^{1/2} h^{q+1/2} v_n \to 0$.
Then, 
for an arbitrary $t$, $\wh \mu_{\gsiraipwcf}^{(t)} (t)$ in \eqref{eq:gsir-aipw-estimator-1} is a consistent estimator for $\mu(t)$. Moreover, for any $t$, 
\be\label{eq:gsir-aipw-asymp-1}
&& n^{1/2}h^{1/2}\{\wh \mu_{\gsiraipwcf}^{(t)}(t) - \mu(t)\} \n\\
&=& n^{-1/2}h^{1/2}\sumi \left[ m(t,Z_i) + \frac{K_h(T_i-t)}{f_{T|Z}(t|Z_i)}\{Y_i - m(t, Z_i)\} - \mu(t) \right] + o_p(1).
\ee
\end{theorem}

In Theorem \ref{thm:gsir-aipw-1}, the rate conditions $u_n \to 0$ and $v_n \to 0$ require the consistency of the estimators of the nuisance parameters, and the condition $n^{1/2} h^{1/2} u_n v_n \to 0$ is the widely-used rate double-robustness condition (see, for example, Assumption 2.4(c) of \cite{colangelo2025double}). Note that, in binary treatment case, it is usually sufficient to assume $n^{1/2} u_n v_n \to 0$, and a commonly used special case is to assume $u_n,v_n = o(n^{-1/4})$ (see, for example, Definition 2 of \cite{smucler2019unifying}, equation 3.8 of \cite{chernozhukov2018double}). Also, Theorem \ref{thm:gsir-aipw-1} additionally requires $n^{1/2} h^{q+1/2} u_n \to 0$ and $n^{1/2} h^{q+1/2} v_n \to 0$ because of the additional smoothing step over $t$ in the continuous treatment case. The next corollary gives the asymptotic normality of $\wh \mu_{ \gsiraipwcf}^{(t)} (t)$ in \eqref{eq:gsir-aipw-estimator-1}, which is based on the asymptotic expansion \eqref{eq:gsir-aipw-asymp-1}.
\begin{corollary}\label{cor:gsir-aipw-1}
    Under the same assumptions as in Theorem \ref{thm:gsir-aipw-1}, and under the additional assumptions that the three functions
    \bse
    E[\{Y(T)- m(t,Z)\}^2 |T = t, Z = z], \quad \frac{\partial^q \{m(t,Z)f_{T,Z}(t,Z)\}}{\partial t^q}, \quad\text{and}\quad  \frac{\partial^q f_{T, Z}(t, Z)}{\partial t^q}
    \ese
    are Lipschitz continuous in $t$ and $\int  |K_T(u)|^3 du <\infty$, we have
    \bse
    n^{1/2}h^{1/2}\{\wh \mu_{\gsiraipwcf}^{(t)} (t) - \mu(t) - h^q B_t\} \xrightarrow{d} N(0, V_t),
    \ese
    where 
    \bse
    B_t =  (q!)^{-1} \int K_T(u)u^q du E\left[ \frac{1}{f_{T, Z}(t, Z)} \frac{\partial^q \{m(t,Z)f_{T,Z}(t,Z)\}}{\partial t^q}  - \frac{m(t,Z)}{f_{T, Z}(t, Z)} \frac{\partial^q f_{T, Z}(t, Z)}{\partial t^q}   \right]
    \ese
    and 
    \bse
    V_t = \int K_T^2(u) du E\left(\frac{1}{f_{T|Z}(t|Z)}E[\{Y(T)-m(t,Z)\}^2 |T = t, Z]\right).
    \ese
\end{corollary}

We also propose a similar analysis of $\wh  \mu_{\gsiraipwcf}^{(T)} (t)$ in \eqref{eq:gsir-aipw-estimator-2}, where an additional bias term will be needed.
\begin{theorem} \label{thm:gsir-aipw-2}
Suppose that all assumptions in Theorem \ref{thm:gsir-aipw-1} hold.
Then, 
for an arbitrary $t$, $\wh \mu_{\gsiraipwcf}^{(T)} (t)$ in \eqref{eq:gsir-aipw-estimator-2} is a consistent estimator for $\mu(t)$. Moreover, for any $t$, 
\be\label{eq:gsir-aipw-asymp-2}
&& n^{1/2}h^{1/2}\{\wh \mu_{\gsiraipwcf}^{(T)}(t) - \mu(t) - C_t\} \n\\
&=& n^{-1/2}h^{1/2}\sumi \left[ m(t,Z_i) + \frac{K_h(T_i-t)}{f_{T|Z}(T_i|Z_i)}\{Y_i - m(T_i, Z_i)\} - \mu(t) \right] + o_p(1),
\ee
 where
\be\label{eq:ct-bias}
C_t = \frac{1}{L} \sum_{l=1}^L E[\{\wh m_l(t,\wh Z_{l}) - m(t,Z)\} - K_h(t-T)q(T,Z)\{\wh m_l(T, \wh Z_{l}) - m(T,Z)\} |O_l^c].
\ee
Here,  $\wh Z_l = \wh z_l (X)$, and $O_l^c$ denotes the observations not in $I_l$ for $l = 1,\ldots, L$.
\end{theorem}

Different from $\wh\mu_{\gsiraipwcf}^{(t)}(t)$ where its bias term $h^q B_t$ is hidden in each summand of the asymptotic expansion \eqref{eq:gsir-aipw-asymp-1}, the bias of $\wh\mu_{\gsiraipwcf}^{(T)}(t)$ is explicitly given as $C_t$ in \eqref{eq:ct-bias}. Specifically, as is shown in Corollary \ref{cor:gsir-aipw-1}, each summand in \eqref{eq:gsir-aipw-asymp-1} does not have mean zero and its expectation instead contains the bias term $h^qB_t$. However, as will be shown in Corollary \ref{cor:gsir-aipw-2} below, each summand in \eqref{eq:gsir-aipw-asymp-2} exactly has mean zero, which contributes to no additional bias term for $\wh\mu_{\gsiraipwcf}^{(T)}(t)$. 
Further analysis of $C_t$ will require the asymptotic expansion of $\wh m_l(t,z)$, and we will later give some discussions in a special case when $\wh m_l(t,z)$ is estimated by the kernel estimators in \eqref{eq:m-f-kernel-est}.

\begin{corollary}\label{cor:gsir-aipw-2}
    Under the same assumptions as in Theorem \ref{thm:gsir-aipw-2}, and under the additional assumptions that the function
    \bse
    E[\{Y(T)- m(t,Z)\}^2 |T = t, Z = z]
    \ese
    are Lipschitz continuous in $t$ and $\int  |K_T(u)|^3 du <\infty$, we have
    \bse
    n^{1/2}h^{1/2}\{\wh \mu_{\gsiraipwcf}^{(T)}(t) - \mu(t) - C_t\} \xrightarrow{d} N(0, V_t),
    \ese
    where $V_t$ is defined in Corollary \ref{cor:gsir-aipw-1} and $C_t$ is given in \eqref{eq:ct-bias}.
\end{corollary}

\subsection{Rate Conditions under Kernel Estimation}\label{sec:discussion-gsir-aipw}
Theorem \ref{thm:gsir-aipw-1} allows any nuisance function estimators of $m(t,z)$ and $q(t,z)$ that satisfy its high-level rate conditions.
Recalling $q(t,z) = 1/f_{T|Z}(t|z)$, we now consider kernel estimation of $m(t,z)$ and $f_{T|Z}(t|z)$ and derive conditions ensuring these requirements hold.
Using the observations in $J_{l,2}$, define
\be\label{eq:m-f-kernel-est}
\wh m_l(t, z) = \frac{\sum_{i \in  J_{l,2}} Y_i K_{\wt h}(t - T_i)K_b(z - \wh Z_{li})}{\sum_{i \in  J_{l,2}} K_{\wt h}(t - T_i)K_b(z - \wh Z_{li})}, \quad
\text{and} \quad 
\wh f_{l, T|Z}(t|z) = \frac{\wh{f}_{l, T,Z}(t, z)}{\wh f_{l,Z}(z)},
\ee
where
\be\label{eq:f-kernel-est-2}
\wh{f}_{l, T,Z}(t, z) = \frac{1}{n_{l,2}}  \sum_{i \in  J_{l,2}} K_{\wt h}(t - T_i) K_b(z - \wh{Z}_{li}), \qquad
\wh f_{l,Z}(z) = \frac{1}{n_{l,2}} \sum_{i \in  J_{l,2}} K_b(z - \wh Z_{li}).
\ee
Note that, regarding the kernel on $T$, we use the notation $K_h$ in \eqref{eq:gsir-aipw-estimator-1} and \eqref{eq:gsir-aipw-estimator-2} and use $K_{\wt h}$ in \eqref{eq:m-f-kernel-est} and \eqref{eq:f-kernel-est-2}. We refer to the kernel $K_h$ appearing in \eqref{eq:gsir-aipw-estimator-1} and \eqref{eq:gsir-aipw-estimator-2} as the localization kernel,  and that in \eqref{eq:m-f-kernel-est} and \eqref{eq:f-kernel-est-2} as the kernel for nuisance parameter estimation. Throughout this paper, for simplicity, we assume the same form and order for both kernels, but allow different bandwidths subject to different rate conditions. More generally, their forms and orders may also be different.

The following lemma gives convergence rates for the kernel estimators in \eqref{eq:m-f-kernel-est}, which can be used as plug-in nuisance function estimators for the GSIR-AIPW estimators.
\begin{lemma}\label{lem:m-f-rate-split}
Suppose that kernel estimators \eqref{eq:m-f-kernel-est}  and \eqref{eq:f-kernel-est-2} are used for $m(t,z)$ and $f_{T|Z}(t|z)$.
If Assumptions  \ref{ass:fZ-bound}--\ref{ass:diff-bound} hold and Assumption \ref{ass:rate} holds for $b$ and $\wt h$, then the rates in \eqref{eq:m-q-vn} are given as
\be\label{eq:un-vn}
u_n = v_n = r_nt_{n,\wt h,b}+r_n b^{-1} + r_n^2 b^{-d-2}\wt h^{-1} + b^s + \wt h^q + (n\wt hb^d)^{-1/2}(\log n)^{1/2},
\ee
and $t_{n,\wt h,b}$ is given in \eqref{eq:t-nhb} .
\end{lemma}
The following corollary gives the optimal rates in \eqref{eq:un-vn} over the bandwidths $\wt h$ and $b$ for a given GSIR rate $r_n$.
\begin{corollary} \label{cor:optimal-aipw-rate}
    Suppose that the kernel bandwidths have the forms $\wt h \asymp n^{-\gamma}$ and $b \asymp n^{-\eta}$ for some $\gamma, \eta > 0$. Also, assume that the GSIR convergence rate is $r_n \asymp n^{-\delta}$, where $0 < \delta \leq 1/2$ is fixed. Then, the optimal convergence rate of $u_n$ and $v_n$ given in \eqref{eq:un-vn} is $n^\zeta$, where 
    \be\label{eq:zeta}
    \zeta = \max\left\{-\frac{s\delta}{s+1}, -\frac{2sq\delta}{s+q(d+s+2)}, -\frac{sq}{s+q(d+2s)}\right\}.
    \ee
\end{corollary}

For $d \in \{1,2\}$ and different kernel orders $(s,q)$, Table \ref{tab:gsir-rate}  reports the admissible range of $\delta$, where $r_n \asymp n^{-\delta}$, such that the resulting nuisance rates satisfy the conditions of Theorem \ref{thm:gsir-aipw-1}. We choose he bandwidth $h$ of the localization kernel to minimize the asymptotic mean square error of the estimator, which leads to $h \asymp n^{-1/(2q+1)}$. Under this choice of the localization bandwidth, as long as $u_n \to 0$ and $v_n \to 0$, the conditions $n^{1/2}h^{q+1/2}u_n\to 0$ and $n^{1/2}h^{q+1/2}v_n\to 0$  hold automatically. Therefore, it suffices to check whether $n^{1/2}h^{1/2}u_nv_n \to 0$ holds, which is equivalent to checking whether $\zeta< - q/\{2(2q+1)\}$.

 Table \ref{tab:gsir-rate} highlights the importance of the convergence rate of the GSIR estimator. When $\delta = 1/4$, corresponding to the optimal rate established by \cite{li2017nonlinear}, the required rate conditions fail to hold in many of the settings considered. In contrast, when $\delta = 1/3$, corresponding to the limiting optimal convergence rate established by \cite{choi2026sharpened}, the rate conditions are satisfied in most settings. Although improving the convergence rate of GSIR is not the focus of this paper, these findings illustrate that such improvements can substantially broaden the range of settings in which the proposed causal inference procedure is theoretically justified.

\begin{lemma}\label{lem:dm-rate-split}
Suppose that kernel estimator \eqref{eq:m-f-kernel-est} is used for $m(t,z)$, and Assumptions  \ref{ass:fZ-bound}--\ref{ass:rate} hold. Further assume that Assumption \ref{ass:additional-T} holds for some $\nu \in \N$. For each $j = 0,1,\ldots, \nu$, let 
\be\label{eq:u-n-nu}
u_{n,j}
&=& r_n\wt h^{-j}(n\wt hb^{d+1})^{-1} \{\log(1/b) + \log(1/\wt h)\}
+
r_n\wt h^{-j}(n\wt hb^{d+2})^{-1/2}  \sqrt{\log(1/b) + \log(1/\wt h)}\n\\
&&
+
r_nb^{-1}\wt h^{-j}
+
r_n^2b^{-d-2}\wt h^{-j-1}
+
b^s
+
\wt h^q
+
(n\wt h^{2j+1}b^d)^{-1/2}
(\log n)^{1/2}.
\ee
 If $u_{n,\nu} = O(1)$, then we have
\bse
\sup_{t, z, \wh z} \left|\frac{\partial^\nu}{\partial t^\nu} \{\wh m_l(t, \wh z) - m(t, z) \}\right| = O_p(u_{n,\nu}).
\ese
\end{lemma}

\begin{corollary}\label{cor:ct-rate}
Suppose that all assumptions in Lemma \ref{lem:dm-rate-split} are satisfied, and let $C_t$ be defined as \eqref{eq:ct-bias}. Then, $C_t = O_p(h^{\wt\nu})$, where $\wt\nu = \nu \wedge q$.
\end{corollary}

In particular, if $\nu \ge q$,  Corollary \ref{cor:ct-rate} implies that $C_t = O(h^q)$. Thus, $\wh\mu_{\gsiraipwcf}^{(T)}(t)$ also has an asymptotic bias of order $O(h^q)$, matching the order of the bias $h^q B_t$ for $\wh\mu_{\gsiraipwcf}^{(t)}(t)$ in Corollary \ref{cor:gsir-aipw-1}. Therefore, the bandwidth of the localization kernel, $h$, can still be chosen as $h \asymp n^{-1/(2q+1)}$, and the admissible requirements for $\delta$ as in Table \ref{tab:gsir-rate} also holds in this case.

\subsection{Variance Implications of Sufficient Confounders}
\label{sec:variance-implications}

Let $W$ be either $W = X$ or $Z = z(X)$ satisfying \eqref{eq:low-dim-ignorability}. By Corollaries \ref{cor:gsir-aipw-1} and \ref{cor:gsir-aipw-2} and Theorem \ref{thm:aipw-1}, the asymptotic variances of kernel-based AIPW estimators, scaled by $\sqrt{nh}$, are of the form $R(K_T)V_W(t)$.   Here, $R(K_T) =\int K_T^2(u)du$ is a constant depending only on the kernel function $K_T$, and 
\bse
V_W(t)  = E\left\{ \frac{\sigma^2(t, W)}{f_{T| W}(t | W)} \right\}, 
\quad \text{where} \quad
\sigma^2(t, w) = \var\{Y(T) | T = t, W = w\}.
\ese
Note that we use the same abuse of notation for $\sigma^2(t,w)$ as in $m(t,w)$.
For a common kernel and bandwidth, comparing $V_Z(t)$ and $V_X(t)$ therefore quantifies the change in this leading variance component. This leading variance term also appears in some existing estimators, such as those of \cite{kennedy2017non}, \cite{huang2018joint} and \cite{colangelo2025double}, as long as $W$ serves as a valid confounder and the regularity conditions are satisfied. The following theorem provides a decomposition for an arbitrary sufficient confounder.

\begin{theorem}\label{thm:var-gain}
Suppose Assumptions \ref{ass:consistency}--\ref{ass:ignorability} hold. Let $Z = z(X)$ be a sufficient confounder satisfying \eqref{eq:low-dim-ignorability}. Then for any $t \in \Omega_T$ such that $E\{Y(t)^2\}<\infty$, we have 
\bse
V_Z(t) - V_X(t) = A(t) - B(t) + C(t),
\ese
where 
\be
A(t) &=& E \left[ \frac{f_{T| X}(t|X)\{m(t,X) - m(t,Z)\}^2}{f^2_{T| Z}(t| Z)}\right], \label{eq:a-t}\\
B(t) &=& E \left[ \frac{\sigma^2(t,X)\{f_{T| X}(t| X) - f_{T| Z}(t | Z)\}^2}{f_{T| X}(t|X)f^2_{T| Z}(t| Z)}\right], \label{eq:b-t}\\
C(t) &=& 2E \left[ \frac{\cov\{\sigma^2(t,X), f_{T| X}(t|X) | Z\}}{f_{T| Z}^2(t|Z)}\right].\label{eq:c-t}
\ee
\end{theorem}

The nonnegative terms $A(t)$ and $B(t)$ measure the loss of
outcome-regression information and the discrepancy between the
conditional treatment densities, respectively, while $C(t)$
is a covariance term that can have either sign.
Thus, for a general sufficient confounder satisfying
\eqref{eq:low-dim-ignorability}, the sign of
$V_Z(t)-V_X(t)$ cannot be determined.
 In contrast, the two implementable SDR constructions yield
definite variance orderings.
For a balancing sufficient confounder satisfying \eqref{eq:T-X-sdr}, $B(t)=C(t)=0$, giving
\bse
V_Z(t)-V_X(t)
=
E\left[
\frac{\{m(t,X)-m(t,Z)\}^2}
{f_{T|X}(t|X)}
\right]
\geq 0,
\ese
with equality if and only if \eqref{eq:nuis-equiv-1} holds
almost surely. For a prognostic-balancing sufficient confounder satisfying \eqref{eq:YT-X-sdr}, $A(t)=B(t)=C(t)=0$, so
$V_Z(t)=V_X(t)$.

These results show that the prognostic-balancing sufficient confounder
preserves the full-covariate leading AIPW variance, whereas
the balancing sufficient confounder may increase it. This provides a further justification for using prognostic-balancing sufficient confounder in the AIPW estimators. The trade-off is that preserving additional outcome information may require a prognostic-balancing sufficient confounder of higher dimension than a balancing sufficient confounder, complicating nonparametric nuisance function estimation and making the required convergence rate conditions harder to satisfy. Nevertheless, compared with using the full covariates $X$, the reduction can still substantially relax the required rate conditions, as illustrated in Section \ref{sec:comp-gsiraipw-aipw}.  Finally, the variance comparisons above concern population-level reductions; when $z(X)$ is estimated, additional rate conditions are needed to control its estimation error.

\subsection{Comparisons between GSIR-AIPW and AIPW}\label{sec:comp-gsiraipw-aipw}
Since GSIR-AIPW uses a prognostic-balancing sufficient confounder, the GSIR-AIPW and AIPW estimators share the same asymptotic variance by Theorem \ref{thm:var-gain}. We therefore focus on comparing their rate conditions. The asymptotic properties of the full-covariate AIPW estimators are analyzed in Appendix \ref{app:aipw-X-asymp}, and the counterparts of $\zeta$, $u_n$, and $v_n$ are denoted by $\zeta'$, $u_n'$, and $v_n'$, respectively. Consider $q = 2$, which is the order of the commonly used Gaussian kernel. When the localization bandwidth is chosen as $h \asymp n^{-1/(2q+1)} = n^{-1/5}$, the admissible upper bound for both $\zeta$ and $\zeta'$ is $-q/\{2(2q+1)\} = -1/5$. Thus, the nuisance function estimators are required to converge faster than $n^{-1/5}$, i.e., we need $u_n,v_n = o(n^{-1/5})$ and $u_n',v_n' = o(n^{-1/5})$.

We first compare the orders of the covariate kernels $K_Z$ and $K_X$ required for the rate conditions in Theorems \ref{thm:gsir-aipw-1} and \ref{thm:aipw-1} to hold.
Although higher-order kernels reduce asymptotic bias, they raise two practical concerns.
First, higher-order kernels are typically sign-changing. In kernel density estimation, this may yield negative density estimates; in Nadaraya-Watson regression, negative weights imply that the estimator may no longer be a genuine local average of the observed responses.
Second, beyond these effects of negative weights, their favorable asymptotic properties may not translate into good finite-sample performance \citep{Botev2010Kernel}, with asymptotic improvements becoming evident only for very large sample sizes \citep{marron1994visual}.
These concerns motivate determining the minimum kernel order required by each estimator, given the GSIR convergence rate and the covariate dimensions.

Let $s_Z$ and $s_X$ denote the orders of $K_Z$ and $K_X$, respectively. Suppose that the GSIR convergence rate satisfies $r_n \asymp n^{-\delta}$ with $\delta > 1/5$. It follows from \eqref{eq:zeta} and \eqref{eq:zeta-X} that $\zeta< -1/5$ and $\zeta' < -1/5$ are equivalent to $s_Z > s_{\mathrm{lb}}$ and $s_X > s_{\mathrm{lb}}'$, where
\be\label{eq:s-lb}
s_{\mathrm{lb}} = \max \left( \frac{1}{5\delta-1}, \frac{2d+4}{20\delta-3}, \frac{2d}{5}\right),
\qquad \text{and} \qquad
s_{\mathrm{lb}}' = \frac{2p}{5}.
\ee
These bounds highlight the benefit of dimension reduction. The lower bound for GSIR-AIPW depends on the reduced dimension $d$ and the GSIR convergence rate $\delta$, whereas that for the original AIPW estimator increases linearly with the full covariate dimension $p$.

For a numerical comparison, we evaluate the lower bounds $s_{\mathrm{lb}}$ and $s_{\mathrm{lb}}'$ in \eqref{eq:s-lb} with $q=2$. As in Table \ref{tab:rate-comparison}, we take $p=5,10,15,20,25,30$, $d=1,2$, and $\delta=1/3,7/24,1/4$. The results are reported in Table \ref{tab:s-lb-comparison}, which shows a pattern similar to that in Table \ref{tab:rate-comparison}. In moderate- and high-dimensional settings ($p \ge 10$), applying GSIR before kernel estimation can reduce the required kernel order. By contrast, when the covariate dimension is low ($p=5$), this advantage largely disappears because the original method based on $X$ already requires only a low-order kernel. This suggests that the benefit of GSIR in relaxing kernel order requirements becomes more pronounced as the covariate dimension $p$ increases.

Next, we compare the convergence rates of the nuisance function estimators used in the GSIR-AIPW estimator with those for the original AIPW estimator. Specifically, we compare  $\zeta$ and $\zeta'$ defined in \eqref{eq:zeta} and \eqref{eq:zeta-X} using the same combinations of $\delta$, $p$, $d$, and $q$ as in Table \ref{tab:s-lb-comparison}, and considering kernel orders $s_Z = s_X = s$ with $s=2$ and $s=4$. The results are reported in Table \ref{tab:rate-comparison-2}. 

When $s=2$, the admissible condition on $\zeta$ is satisfied only when $d=1$ and $\delta=1/3$. In contrast, the corresponding admissible condition on $\zeta'$ is not satisfied in any of the settings considered when GSIR is not applied. When $s=4$, the condition on $\zeta$ is satisfied whenever the GSIR convergence rate is not too slow ($\delta=1/3$ or $\delta=7/24$) and the sufficient confounder dimension $d$ is either 1 or 2. By contrast, the condition on $\zeta'$ is only satisfied when when $p=5$. Hence, with the same kernel order, applying GSIR helps achieves the rate conditions. In particular, when the sufficient confounder dimension $d$ is small and GSIR converges fast, as with the improved rate of \cite{choi2026sharpened} that can approach $n^{-1/3}$, GSIR-AIPW satisfies the rate conditions, whereas AIPW fails to satisfy them once the covariate dimension $p$ is moderate.

In conclusion, when the covariate dimension is high, nonparametric estimation based directly on $X$ suffers from the curse of dimensionality. Consequently, the rate double-robustness condition ($n^{1/2} h^{1/2} u_n' v_n' \to 0$) is difficult to satisfy unless strong smoothness conditions are imposed on the nuisance functions and high-order kernels are used. However, applying GSIR in the first step allows the nuisance functions to be estimated using the sufficient confounder. When the sufficient confounder is low-dimensional and GSIR converges sufficiently fast, the condition can be satisfied under substantially weaker smoothness assumptions without inflating the asymptotic variance of the estimator.

\section{Simulation Studies}\label{sec:simulation}

In this section, we conduct simulation studies to evaluate the finite-sample performance of our proposed methods. We generate $X = (X_1,\ldots,X_p)^\top \sim N(0,I_p)$, with $p=10$ for Models M1 and M3 and $p=20$ for Model M2. We consider the following models 
\begin{enumerate}
    \item[M1:] $Z_1 = (\sum_{i=1}^8 X_i^2 - 8)/4$, \quad  $T = 0.6 Z_1 + \epsilon_1$, \quad  $Y(t) = t - 2 Z_1 + \epsilon_2$,

    \item[M2:] $Z_1 = (\sum_{i=1}^8 X_i^2 - 8)/4$, \quad  $T = 0.6 Z_1 + \epsilon_1$, \quad  $Y(t) = t - 2 Z_1 + \epsilon_2$,
    
    \item[M3:] $Z_1 = (X_1^2 - 1)/\sqrt{2}$, \quad $T = 0.6 Z_1 + \epsilon_1$, \quad $Y(t) = t - 2Z_1 + \epsilon_2 $,
    
\end{enumerate}
where $\epsilon_1 \sim N(0,1)$ and $\epsilon_2 \sim N(0, 0.8^2)$ are mutually independent of $X$. In all three models, $Z_1$ has mean zero and unit variance, serving as a sufficient confounder, and the true dose-response curve is $\mu(t) = t$.  Models M1 and M2 share the same treatment and outcome models, except that the dimension of covariates is different. In both models, nonlinear SDR has structural dimension one, while linear SDR requires an eight-dimensional subspace spanned by $(X_1, \ldots, X_8).$ Model M3 represents a simpler setting where both nonlinear and linear SDR have structural dimension one.

We generate 500 simulated data sets for each sample size $n \in \{600, 1200, 1800\}$.  For each simulated data set, we compare IPW, OR, and AIPW estimators constructed using three different covariate representations. The first uses the full-covariate vector $X$ without dimension reduction. The second uses sufficient confounder estimated by the linear SDR method MAVE \citep{xia-tong-li-zhu-2002,xia-2007}, and the third uses sufficient confounder estimated by our proposed nonlinear SDR method, GSIR. For the dimension-reduction approaches, we first estimate the sufficient confounder and then use estimated sufficient confounder to estimate the generalized propensity score and outcome regression functions. We evaluate IPW and OR without cross-fitting and AIPW both with and without cross-fitting. Cross-fitting is implemented using $L=3$ folds following the procedure described in Section \ref{sec:cross-fitting}.  For all kernel-based nuisance function estimators and the localization kernels in the IPW and AIPW estimators, bandwidths are selected using Silverman's rule of thumb \citep{silverman1986density}. That is, for a $d$-dimensional random vector $V = (V_1,\ldots, V_d)^\top$, $h_j = \{4/(d+2)\}^{1/(d+4)} \wh \sigma_j n^{-1/(d+4)}$, $j = 1,\ldots, d$, where $\wh \sigma_j$ is the sample standard deviation of $V_j$, and $n$ is the sample size used to fit the corresponding estimator.
Throughout the simulation study, the structural dimensions of MAVE and GSIR are treated as known. 

We evaluate estimation performance using two metrics, integrated absolute mean bias and integrated root mean squared error (RMSE), which are also used in \cite{kennedy2017non}. Let $\wh \mu_s(t)$ denote the estimated dose-response curve from the $s$th simulation, for $s = 1,\ldots, S$, where $S = 500$. We denote 
\be
 \wh{\text{Bias}} &=& \int_{\ca T^*} \left|S^{-1}\sum_{s=1}^{S} \wh \mu_s  (t) - \mu(t) \right| f_T(t) dt,\label{eq:bias-def}\\
 \wh{\text{RMSE}} &=& \int_{\ca T^*} \left[S^{-1}\sum_{s=1}^{S} \{\wh \mu_s  (t) - \mu(t)\}^2 \right]^{1/2} f_T(t) dt.\label{eq:rmse-def}
\ee
Here, $\ca T^*$ is a trimmed treatment region obtained by excluding 10\% of the probability mass from each tail of the marginal distribution of $T$, and $ f_T(t)$ denotes the marginal density of $T$. In implementation, these integrals are approximated using Riemann sums over an equally spaced grid of 50 treatment values in $\ca T^*$.

The simulation results for integrated absolute mean bias and integrated root mean squared error are reported in Tables \ref{tb:simulation_bias} and \ref{tb:simulation_rmse}, respectively. Overall, the GSIR-based estimators perform better than the estimators based on the full-covariate vector. However, in some settings in Model M3, the cross-fitted AIPW estimator based on the full-covariate vector has smaller integrated absolute mean bias but larger RMSE than the dimension-reduction-based estimators. The GSIR-based estimators also generally outperform the MAVE-based estimators in Models M1 and M2, as they use a much lower-dimensional sufficient confounder for nuisance parameter estimation. In Model M3, both MAVE and GSIR yield a one-dimensional sufficient confounder, so GSIR has no advantage in the dimension used for nuisance estimation. Their relative performance is more mixed, with MAVE achieving lower RMSE for the cross-fitted AIPW estimators. These patterns suggest that GSIR is particularly useful when nonlinear SDR permits substantially greater dimension reduction than linear SDR.

Among the proposed methods, GSIR-AIPW without cross-fitting generally achieves smaller RMSE than GSIR-OR and GSIR-IPW. GSIR-OR also generally outperforms GSIR-IPW.
The (GSIR-)IPW estimators rely directly on the reciprocal of the estimated generalized propensity score. Small estimated scores can therefore produce large, unstable weights and impair performance. For GSIR-AIPW, however, the cross-fitted version has larger RMSE than its non-cross-fitted counterpart in all settings considered, with larger differences at smaller sample sizes. This may reflect the reduced sample sizes available for estimating the sufficient confounder and nuisance functions under sample splitting. 
Cross-fitting is introduced primarily to facilitate the theoretical analysis underlying asymptotic inference, rather than to guarantee improved finite-sample accuracy.
Motivated by these simulation results, we use the GSIR-AIPW estimators without cross-fitting to estimate the dose-response curve in the real-data analysis.

\begin{table}[!ht]
\centering
\footnotesize
\begin{tabular*}{\textwidth}
{@{\extracolsep{\fill}}cc*{9}{c}@{}}
\hline\hline
\multicolumn{11}{c}{Panel A: IPW and OR estimators} \\
\hline
& & \multicolumn{3}{c}{IPW$(T)$}
& \multicolumn{3}{c}{IPW$(t)$}
& \multicolumn{3}{c}{OR} \\
\cline{3-5}\cline{6-8}\cline{9-11}
Model & $n$ 
& Full $X$ & MAVE & GSIR
& Full $X$ & MAVE & GSIR
& Full $X$ & MAVE & GSIR \\
\hline
\multirow{3}{*}{M1}
& \multirow{1}{*}{600} 
& $0.485$ & $0.485$ & $0.288$
& $0.481$ & $0.478$ & $0.230$
& $0.419$ & $0.386$ & $0.099$ \\
& \multirow{1}{*}{1200} 
& $0.487$ & $ 0.487$ & $0.224$
& $0.483$ & $0.480$ & $0.181$
& $0.401$ & $0.363$ & $0.089$ \\
& \multirow{1}{*}{1800} 
& $0.486$ & $0.486$ & $0.191$
& $0.483$ & $0.480$ & $0.156$
& $0.392$ & $0.353$ & $0.081$ \\
\hline
\multirow{3}{*}{M2}
& \multirow{1}{*}{600} 
& $0.483$ & $0.485$ & $0.340$
& $0.480$ & $0.480 $ & $0.285$
& $0.514$ & $0.413 $ & $0.115$ \\
& \multirow{1}{*}{1200} 
& $0.481$ & $0.484$ & $0.304$
& $ 0.479 $ & $0.479$ & $0.241$
& $0.513$ & $0.384$ & $ 0.066$ \\
& \multirow{1}{*}{1800} 
& $0.480 $ & $0.484$ & $ 0.241 $
& $0.479$ & $0.478$ & $0.184$
& $0.511$ & $0.366$ & $0.057$ \\
\hline

\multirow{3}{*}{M3}
& \multirow{1}{*}{600} 
& $0.432$ & $0.249$ & $0.255$
& $0.427$ & $0.228$ & $0.216$
& $0.413$ & $0.213$ & $0.103$ \\
& \multirow{1}{*}{1200} 
& $0.433$ & $0.226$ & $0.205$
& $0.429$ & $0.208$ & $0.171$
& $0.394$ & $0.185$ & $0.099$ \\
& \multirow{1}{*}{1800} 
& $0.433$ & $0.219$ & $0.190$
& $0.429$ & $0.202$ & $0.158$
& $0.383$ & $0.172$ & $0.096$ \\
\hline\hline
\end{tabular*}

\vspace{6pt}

\begin{tabular*}{\textwidth}
{@{\extracolsep{\fill}}ccc*{6}{c}@{}}
\hline\hline
\multicolumn{9}{c}{Panel B: AIPW estimators} \\
\hline
& & & \multicolumn{3}{c}{AIPW$(T)$}
& \multicolumn{3}{c}{AIPW$(t)$} \\
\cline{4-6}\cline{7-9}
Model & $n$ & $L$
& Full $X$ & MAVE & GSIR
& Full $X$ & MAVE & GSIR \\
\hline
\multirow{6}{*}{M1}
& \multirow{2}{*}{600} & 1
& $0.429$ & $0.396$ & $0.095$
& $0.429$ & $0.395$ & $0.102$ \\
& & 3
& $0.405$ & $0.396$ & $0.089$
& $0.394$ & $0.383$ & $0.118$ \\
& \multirow{2}{*}{1200} & 1
& $0.401$ & $0.361$ & $0.022$
& $0.401$ & $0.360$ & $0.042$ \\
& & 3
& $ 0.382$ & $0.375$ & $ 0.073$
& $ 0.375$ & $0.369$ & $0.099$ \\
& \multirow{2}{*}{1800} & 1
& $0.391$ & $0.351$ & $0.018$
& $0.390$ & $0.350$ & $0.038$ \\
& & 3
& $0.358$ & $0.356$ & $0.056$
& $0.354$ & $0.351$ & $0.080$ \\
\hline

\multirow{6}{*}{M2}
& \multirow{2}{*}{600} & 1
& $0.514 $ & $0.410 $ & $0.030 $
& $0.514 $ & $0.410$ & $0.067 $ \\
& & 3
& $1.011$ & $0.374$ & $0.288$
& $1.005$ & $0.376$ & $0.304$ \\
& \multirow{2}{*}{1200} & 1
& $0.517 $ & $ 0.388 $ & $0.072$
& $ 0.517$ & $0.388$ & $0.069$ \\
& & 3
& $0.853$ & $0.346$ & $0.186$
& $0.853$ & $0.348$ & $0.206$ \\
& \multirow{2}{*}{1800} & 1
& $0.511$ & $0.362$ & $0.020$
& $0.511$ & $0.362$ & $0.011$ \\
& & 3
& $0.811$ & $0.349$ & $0.101$
& $0.810$ & $0.347$ & $0.123$ \\
\hline

\multirow{6}{*}{M3}
& \multirow{2}{*}{600} & 1
& $0.412$ & $0.120$ & $0.041$
& $0.412$ & $0.128$ & $0.059$ \\
& & 3
& $0.084$ & $0.127$ & $0.128$
& $0.080$ & $0.146$ & $0.157$ \\
& \multirow{2}{*}{1200} & 1
& $0.393$ & $0.099$ & $0.044$
& $0.393$ & $0.105$ & $0.057$ \\
& & 3
& $0.065$ & $0.103$ & $0.085$
& $0.059$ & $0.117$ & $0.111$ \\
& \multirow{2}{*}{1800} & 1
& $0.382$ & $0.092$ & $0.046$
& $0.382$ & $0.096$ & $0.055$ \\
& & 3
& $0.077$ & $0.096$ & $0.072$
& $0.070$ & $0.105$ & $0.096$ \\
\hline\hline
\end{tabular*}
\caption{ Bias for five estimators, IPW$(T)$, IPW$(t)$, OR, AIPW$(T)$, AIPW$(t)$, which correspond to the estimators identified by \eqref{eq:ipw-ident-2}, \eqref{eq:ipw-ident-1}, \eqref{eq:or-ident}, \eqref{eq:aipw-ident-2}, \eqref{eq:aipw-ident-1}, respectively. Each estimator involves three covariate representations: using full $X$, linear SDR by MAVE, and nonlinear SDR by GSIR. Within AIPW$(T)$ and AIPW$(t)$, estimators without cross-fitting ($L=1$) and with cross-fitting ($L=3$) are considered. The entries are estimation biases calculated using \eqref{eq:bias-def}.}
\label{tb:simulation_bias}
\end{table}

\begin{table}[!ht]
\centering

\begin{tabular*}{\textwidth}
{@{\extracolsep{\fill}}cc*{9}{c}@{}}
\hline\hline
\multicolumn{11}{c}{Panel A: IPW and OR estimators} \\
\hline
& & \multicolumn{3}{c}{IPW$(T)$}
& \multicolumn{3}{c}{IPW$(t)$}
& \multicolumn{3}{c}{OR} \\
\cline{3-5}\cline{6-8}\cline{9-11}
Model & $n$ 
& Full $X$ & MAVE & GSIR
& Full $X$ & MAVE & GSIR
& Full $X$ & MAVE & GSIR \\
\hline

\multirow{3}{*}{M1}
& \multirow{1}{*}{600} 
& $0.492$ & $0.492$ & $0.319$
& $0.489$ & $0.486$ & $0.268$
& $0.430$ & $0.398$ & $0.134$ \\
& \multirow{1}{*}{1200} 
& $0.491$ & $ 0.491$ & $0.249$
& $0.488$ & $0.485$ & $0.210$
& $0.407$ & $0.370$ & $0.111$ \\
& \multirow{1}{*}{1800} 
& $0.489$ & $0.489$ & $0.212$
& $0.486$ & $0.483$ & $0.178$
& $0.396$ & $0.359$ & $0.099$ \\
\hline

\multirow{3}{*}{M2}
& \multirow{1}{*}{600} 
& $0.492$ & $0.492$ & $0.371$
& $0.490$ & $0.488$ & $0.321$
& $ 0.514$ & $0.413$ & $0.115$ \\
& \multirow{1}{*}{1200} 
& $0.486$ & $0.488$ & $0.326$
& $0.485$ & $0.483$ & $0.269$
& $0.517$ & $0.391$ & $0.094$ \\
& \multirow{1}{*}{1800} 
& $0.484$ & $0.487$ & $0.264$
& $0.483$ & $0.482$ & $0.212$
& $0.514$ & $0.371$ & $0.079$ \\
\hline

\multirow{3}{*}{M3}
& \multirow{1}{*}{600} 
& $0.438$ & $0.271$ & $0.282$
& $0.434$ & $0.250$ & $0.246$
& $0.422$ & $0.232$ & $0.135$ \\

& \multirow{1}{*}{1200} 
& $0.436$ & $0.241$ & $0.225$
& $0.432$ & $0.223$ & $0.194$
& $0.399$ & $0.198$ & $0.120$ \\
& \multirow{1}{*}{1800} 
& $0.435$ & $0.230$ & $0.206$
& $0.432$ & $0.213$ & $0.176$
& $0.387$ & $0.182$ & $0.112$ \\
\hline\hline
\end{tabular*}

\vspace{6pt}

\begin{tabular*}{\textwidth}
{@{\extracolsep{\fill}}ccc*{6}{c}@{}}
\hline\hline
\multicolumn{9}{c}{Panel B: AIPW estimators}\\
\hline
& & & \multicolumn{3}{c}{AIPW$(T)$}
& \multicolumn{3}{c}{AIPW$(t)$} \\
\cline{4-6}\cline{7-9}
Model & $n$ & $L$
& Full $X$ & MAVE & GSIR
& Full $X$ & MAVE & GSIR \\
\hline

\multirow{6}{*}{M1}
& \multirow{2}{*}{600} & 1
& $0.429$ & $0.396$ & $0.095$
& $0.429$ & $0.395$ & $0.102$ \\
& & 3
& $0.565$ & $0.559$ & $0.179$
& $0.556$ & $0.546$ & $0.196$ \\
& \multirow{2}{*}{1200} & 1
& $0.406$ & $0.368$ & $0.070$
& $0.406$ & $0.367$ & $0.078$ \\
& & 3
& $ 0.487$ & $0.478$ & $0.124$
& $ 0.483$ & $0.472$ & $0.141$ \\
& \multirow{2}{*}{1800} & 1
& $0.395$ & $0.356$ & $0.059$
& $0.395$ & $0.355$ & $0.068$ \\
& & 3
& $0.434$ & $0.431$ & $0.099$
& $0.431$ & $0.426$ & $0.115$ \\
\hline

\multirow{6}{*}{M2}
& \multirow{2}{*}{600} & 1
& $0.522$ & $0.422$ & $ 0.135$
& $ 0.522$ & $0.422$ & $0.142$ \\
& & 3
& $1.292$ & $0.529$ & $ 0.339$
& $ 1.286$ & $0.523$ & $0.351$ \\
& \multirow{2}{*}{1200} & 1
& $0.517$ & $0.388$ & $0.072$
& $0.517$ & $0.388$ & $0.069$ \\
& & 3
& $1.022$ & $0.440$ & $0.226$
& $1.022$ & $0.439$ & $0.242$ \\
& \multirow{2}{*}{1800} & 1
& $0.514$ & $0.367$ & $0.060$
& $0.514$ & $0.367$ & $0.056$ \\
& & 3
& $0.932$ & $0.417$ & $0.134$
& $0.932$ & $0.414$ & $0.151$ \\
\hline

\multirow{6}{*}{M3}
& \multirow{2}{*}{600} & 1
& $0.421$ & $0.153$ & $0.095$
& $0.421$ & $0.159$ & $0.105$ \\
& & 3
& $0.298$ & $0.170$ & $0.231$
& $0.287$ & $0.184$ & $0.253$ \\
& \multirow{2}{*}{1200} & 1
& $0.398$ & $0.122$ & $0.079$
& $0.398$ & $0.127$ & $0.089$ \\
& & 3
& $0.222$ & $0.132$ & $0.150$
& $0.216$ & $0.141$ & $0.167$ \\
& \multirow{2}{*}{1800} & 1
& $0.386$ & $0.109$ & $0.072$
& $0.386$ & $0.113$ & $0.080$ \\
& & 3
& $0.180$ & $0.117$ & $0.122$
& $0.176$ & $0.124$ & $0.138$ \\
\hline\hline
\end{tabular*}
\caption{ RMSE for five estimators in the simulation studies. The meanings of the columns are same as those in Table \ref{tb:simulation_bias}, and the entries are the RMSEs calculated using \eqref{eq:rmse-def}.}
\label{tb:simulation_rmse}
\end{table}

\section{ Data Application}\label{sec:application}

We illustrate the proposed method using data from the Job Corps program, a large-scale, publicly funded job training program implemented in the United States in the mid-1990s to support disadvantaged youth. The Job Corps data have been analyzed in \cite{flores2012estimating}, \cite{huber2020direct}, and \cite{colangelo2025double}; we refer readers to these studies for details on the program, study design, and data summaries. We use the cleaned data set of \cite{huber2020direct}, restricting attention to the 3,906 individuals who completed at least 40 hours of training and retaining the 40 baseline covariates used in that study. We assume that the identifying conditions required for causal estimation hold, as their plausibility was discussed in \cite{flores2012estimating}; see also \cite{colangelo2025double}. 

In this application, the continuous treatment $T$ is the total number of hours of academic and vocational training received during the first year after assignment, and the outcome $Y(T)$ is the proportion of weeks employed during the second year after assignment. Let $X$ denote the vector of 40 baseline covariates. We estimate the average dose-response function $\mu(t) = E\{Y(t)\}$ to assess how employment outcomes vary with training length. To reduce the dimension of the covariates, we apply GSIR to $\{Y(T),T\}$ versus $X$ and find a one-dimensional sufficient confounder. 
We then implement the proposed GSIR-AIPW estimator without cross-fitting. For the kernel-based nuisance function estimators, bandwidths are selected using Silverman’s rule of thumb. 

\begin{figure}[ht]
\centering
\includegraphics[width=0.8\textwidth]{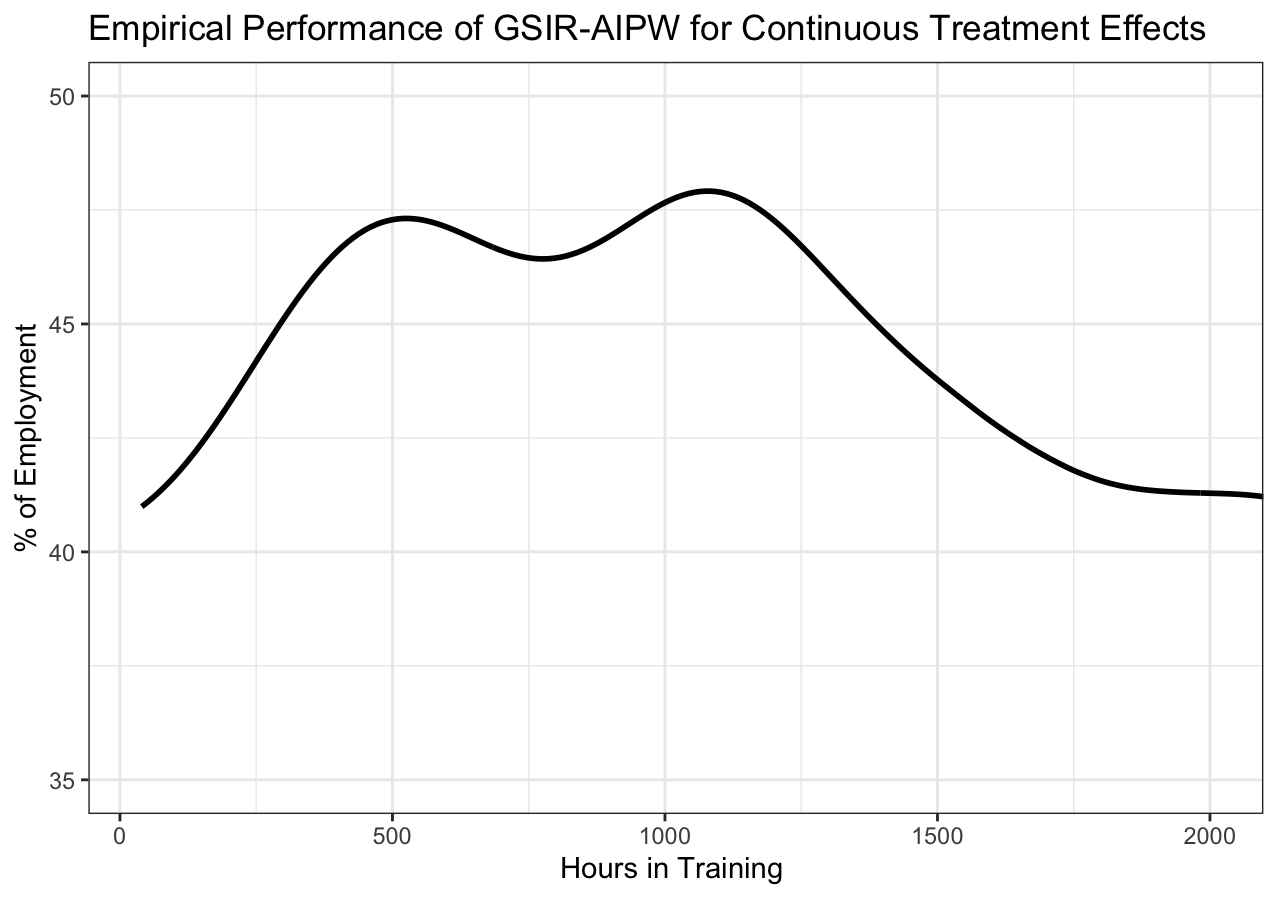}
\caption{Estimated average dose-response function for the Job Corps data obtained using the proposed GSIR-AIPW estimator with a one-dimensional sufficient confounder.}
\label{fig:realdata}
\end{figure}

Figure \ref{fig:realdata} presents the estimated average dose-response function for $t \in [40, 2000]$. The average employment rate initially increases, reaches a local maximum at around 500 hours, then declines slightly before rising to a second and higher peak at approximately $t = 1083$ hours. It then gradually decreases at higher treatment levels. This inverted-W relationship between training duration and employment is also reported in \cite{colangelo2025double}. The results suggest that approximately 500 to 1100 hours of training is associated with the highest proportion of weeks employed during the second year after assignment.

\section{Discussion}\label{sec:discussion}

Several directions warrant further investigation. First, it would be also valuable to construct prognostic sufficient confounder, defined by \eqref{eq:prognostic}, without requiring the balancing property. By Theorem \ref{thm:var-gain}, such a $z(X)$ satisfies $A(t)=C(t)=0$, giving
\bse
V_Z(t)-V_X(t)=-B(t)\leq 0,
\ese
with strict inequality whenever
\bse
P\left\{
\sigma^2(t,X)>0,\,
f_{T|X}(t|X)\neq f_{T|Z}(t|Z)
\right\}>0.
\ese
Thus, removing treatment-assignment information that is irrelevant
to the outcome can reduce the leading AIPW variance.
This is closely related to the efficiency gains from excluding
overadjustment variables in nonparametric causal effect estimation
\citep{rotnitzky2020efficient}.

However, directly targeting \eqref{eq:prognostic} is not easy, because it involves the potential outcomes $Y(t)$, which are only partially observed, so it falls outside classical SDR. 
An interesting observation is that, under Assumptions \ref{ass:consistency}--\ref{ass:ignorability}, \eqref{eq:prognostic} is equivalent to 
\bse
Y(T) \indep X | \{T, z(X)\},
\ese
as shown in Proposition \ref{prop:partial-sdr} in the Supplement.
This suggests a partial SDR approach
\citep{chiaromonte-cook-li-2002}.
However, existing partial SDR methods primarily consider linear
reductions \citep{wen2007optimal,feng2013partial,li2023variable},
while nonlinear partial SDR with a continuous conditioning variable
remains largely unexplored; see \cite[Section 4.5.6]{jones2023dimension}
for related work in the case of categorical conditioning variables. We leave this extension for future research.

Second, although the present work focuses on generalized sliced inverse regression (GSIR), the proposed framework is applicable to a broad class of nonlinear SDR methods, including generalized sliced average variance estimation (GSAVE), other kernel-based SDR methods, and more recent deep learning approaches. Developing unified theoretical guarantees for causal inference based on a wider range of SDR estimators remains an important direction for future research. 

Third, while our theoretical results focus on a single treatments, extending the framework to longitudinal treatments or dynamic treatment regimes would substantially broaden its applicability.

 Another practical consideration is the selection of the structural dimension. Throughout the theoretical development, the structural dimensions of both nonlinear and linear SDR are assumed to be known. In practice, however, these dimensions must be estimated from the data. Fortunately, the structural dimension is often quite small in many applications, particularly for nonlinear SDR, where one or two sufficient predictors frequently capture the relevant information. A simple diagnostic is to examine scatter plots of the treatment or outcome against the estimated sufficient predictors under different candidate dimensions. Once the true structural dimension has been reached, additional estimated sufficient predictors would be expected to exhibit little systematic relationship with either the treatment or the outcome. More formal procedures are also available. For nonlinear SDR, the cross-validated BIC criteria proposed by \cite{li-artemiou-li-2011} and \cite{li2017nonlinear} provide consistent estimators of the structural dimension. For linear SDR, several dimension selection procedures have been developed and are summarized in \cite[Chapters 9--10]{li2018sufficient}, while dimension selection for MAVE is implemented in the software package described by \cite{Xia2025mave}. Incorporating uncertainty from structural dimension selection into the asymptotic analysis is an interesting topic for future research.

Overall, the proposed GSIR-based IPW, OR, and AIPW estimators show that sufficient dimension reduction offers a principled approach to mitigating the curse of dimensionality in causal inference with continuous treatments. GSIR finds a low-dimensional sufficient confounder that preserves the ability to identify the dose-response curve. By estimating the nuisance functions based on this sufficient confounder rather than on $X$, the proposed framework makes flexible nonparametric nuisance parameter estimation more easily compatible with existing response-dose estimators. 
When the dimension of the sufficient confounder is small so that GSIR converges sufficiently fast, this leads to faster convergence rates and weaker rate requirements. We hope that this work motivates further integration of modern sufficient dimension reduction techniques with semiparametric causal inference methods.

\bibliographystyle{agsm}
\bibliography{reference}

@article{hansen2008prognostic,
  title={The prognostic analogue of the propensity score},
  author={Hansen, Ben B},
  journal={Biometrika},
  volume={95},
  number={2},
  pages={481--488},
  year={2008},
  publisher={Oxford University Press}
}

@article{li2023variable,
  title={Variable-dependent partial dimension reduction},
  author={Li, Lu and Tan, Kai and Wen, Xuerong Meggie and Yu, Zhou},
  journal={Test},
  volume={32},
  number={2},
  pages={521--541},
  year={2023},
  publisher={Springer}
}

@article{wen2007optimal,
  title={Optimal sufficient dimension reduction in regressions with categorical predictors},
  author={Wen, Xuerong and Cook, R Dennis},
  journal={Journal of Statistical Planning and Inference},
  volume={137},
  number={6},
  pages={1961--1978},
  year={2007},
  publisher={Elsevier}
}

@phdthesis{jones2023dimension,
  title={Dimension reduction for regression: Theoretical and methodological developments},
  author={Jones, Benjamin},
  year={2023},
  school={Cardiff University}
}

@article{feng2013partial,
  title={On partial sufficient dimension reduction with applications to partially linear multi-index models},
  author={Feng, Zhenghui and Wen, Xuerong Meggie and Yu, Zhou and Zhu, Lixing},
  journal={Journal of the American Statistical Association},
  volume={108},
  number={501},
  pages={237--246},
  year={2013},
  publisher={Taylor \& Francis}
}

@book{li2018sufficient,
  title={Sufficient Dimension Reduction: Methods and Applications with R},
  author={Li, B.},
  isbn={9781498704489},
  series={Chapman \& Hall/CRC Monographs on Statistics and Applied Probability},
  year={2018},
  publisher={CRC Press}
}

@article{li1991sliced,
author = {Li, Ker-Chau},
 journal = {Journal of the American Statistical Association},
 number = {414},
pages = {316--327},
  title = {Sliced inverse regression for dimension reduction},
  volume = {86},
    year = {1991}
}

@article{Cook1991,
author = {Cook, R Dennis and Weisberg, Sanford},
number = {414},
pages = {328--332},
title = {{Sliced Inverse Regression for Dimension Reduction: Comment}},
journal = {Journal of the American Statistical Association},
volume = {86},
year = {1991}
}

@ARTICLE{li2007,
title = {On directional regression for dimension reduction},
author = {Li, Bing and Wang, Shaoli},
year = {2007},
journal = {Journal of the American Statistical Association},
volume = {102},
pages = {997-1008},
url = {}
}

@article{li-1992,
  title={On Principal Hessian Directions for Data Visualization and Dimension Reduction: Another Application of Stein's Lemma},
  author={Li, K.-C.},
  journal={Journal of the American Statistical Association},
  volume={87},
  pages={1025-1039},
  year={1992}
}

@article{xia-tong-li-zhu-2002,
  title={An adaptive estimation of dimension reduction space},
  author={Xia, Y. and Tong, H. and Li, W. K. and Zhu, L.-X.},
  journal={Journal of Royal Statistical Society, Series B},
  volume={64},
  pages={363--410},
  year={2002},
}

@article{li-artemiou-Li-2011,
  title={Principal support vector machines for linear and nonlinear sufficient dimension reduction},
  author={Li, B. and Artemiou, A. and Li, L.},
  journal={The Annals of Statistics},
  volume={39},
  pages={3182--3210},
  year={2011},
}

@article{lee2013,
  title={A general theory for nonlinear sufficient dimension reduction: formulation and estimation},
  author={Lee, Kuang-Yao and Li, Bing and Chiaromonte, Francesca},
  journal={The Annals of Statistics},
  volume={41},
  year={2013},
}

@article{ma2013review,
author = {Ma, Yanyuan and Zhu, Liping},
title = {A Review on Dimension Reduction},
journal = {International Statistical Review},
volume = {81},
number = {1},
pages = {134-150},
doi = {https://doi.org/10.1111/j.1751-5823.2012.00182.x},
year = {2013}
}

@inproceedings{
liang2022nonlinear,
title={Nonlinear Sufficient Dimension Reduction with a Stochastic Neural Network},
author={Siqi Liang and Yan Sun and Faming Liang},
booktitle={Advances in Neural Information Processing Systems},
editor={Alice H. Oh and Alekh Agarwal and Danielle Belgrave and Kyunghyun Cho},
year={2022}
}

@article{chen2024deep,
author = {YinFeng Chen and YuLing Jiao and Rui Qiu and Zhou Yu},
title = {{Deep nonlinear sufficient dimension reduction}},
volume = {52},
journal = {The Annals of Statistics},
number = {3},
publisher = {Institute of Mathematical Statistics},
pages = {1201 -- 1226},
year = {2024},
doi = {10.1214/24-AOS2390}
}

@article{li2017nonlinear,
author = {B. Li and J. Song},
title = {Nonlinear sufficient dimension reduction for functional data},
journal = {The Annals of Statistics},
year = {2017},
pages = {1059--1095},
 }

@Manual{r,
    title = {R: A Language and Environment for Statistical Computing},
    author = {{R Core Team}},
    organization = {R Foundation for Statistical Computing},
    address = {Vienna, Austria},
    year = {2021},
    url = {https://www.R-project.org/},
  }

@article{xia-2007,
  title={A constructive approach to the estimation of dimension reduction directions},
  author={Xia, Y.},
  journal={The Annals of Statistics},
  volume={35},
  pages={2654--2690},
  year={2007},
}

@article{chiaromonte-cook-li-2002,
  title={Sufficient Dimension Reduction in Regressions with Categorical Predictors},
  author={Chiaromonte, F. and Cook, R. D. and Li, B.},
  journal={The Annals of Statistics},
  volume={30},
  pages={475-497},
  year={2002}
}

@article{luo2020matching,
  title={Matching using sufficient dimension reduction for causal inference},
  author={Luo, Wei and Zhu, Yeying},
  journal={Journal of Business \& Economic Statistics},
  volume={38},
  number={4},
  pages={888--900},
  year={2020},
  publisher={Taylor \& Francis}
}

@article{huang2017joint,
  title={Joint sufficient dimension reduction and estimation of conditional and average treatment effects},
  author={Huang, Ming-Yueh and Chan, Kwun Chuen Gary},
  journal={Biometrika},
  volume={104},
  number={3},
  pages={583--596},
  year={2017},
  publisher={Oxford University Press}
}

@article{kennedy2017non,
  title={Non-parametric methods for doubly robust estimation of continuous treatment effects},
  author={Kennedy, Edward H and Ma, Zongming and McHugh, Matthew D and Small, Dylan S},
  journal={Journal of the Royal Statistical Society Series B: Statistical Methodology},
  volume={79},
  number={4},
  pages={1229--1245},
  year={2017},
  publisher={Oxford University Press}
}

@article{huang2018joint,
  title={Joint sufficient dimension reduction for estimating continuous treatment effect functions},
  author={Huang, Ming-Yueh and Chan, Kwun Chuen Gary},
  journal={Journal of multivariate analysis},
  volume={168},
  pages={48--62},
  year={2018},
  publisher={Elsevier}
}

@article{galvao2015uniformly,
  title={Uniformly semiparametric efficient estimation of treatment effects with a continuous treatment},
  author={Galvao, Antonio F and Wang, Liang},
  journal={Journal of the American Statistical Association},
  volume={110},
  number={512},
  pages={1528--1542},
  year={2015},
  publisher={Taylor \& Francis}
}

@article{colangelo2025double,
author = {Kyle Colangelo and Ying-Ying Lee},
title = {Double Debiased Machine Learning Nonparametric Inference with Continuous Treatments},
journal = {Journal of Business \& Economic Statistics},
volume = {0},
number = {0},
pages = {1--13},
year = {2025},
publisher = {ASA Website},
doi = {10.1080/07350015.2025.2505487}
}

@book{tsybakov2009introduction,
author="Tsybakov, Alexandre B.",
title="Introduction to Nonparametric Estimation",
year="2009",
publisher="Springer New York",
address="New York, NY",
doi="10.1007/978-0-387-79052-7_1"
}

@article{tang2025belted,
      title={Belted and Ensembled Neural Network for Linear and Nonlinear Sufficient Dimension Reduction}, 
      author={Yin Tang and Bing Li},
      journal={Journal of American Statistical Association, {\em to appear}},
      year={2025} 
}

@article{li2025relative,
  title={On relative universality, regression operator, and conditional independence},
  author={Li, Bing and Jones, Ben and Artemiou, Andreas},
  journal={arXiv preprint arXiv:2504.11044},
  year={2025}
}

@article{rubin1974estimating,
  title={Estimating causal effects of treatments in randomized and nonrandomized studies.},
  author={Rubin, Donald B},
  journal={Journal of educational Psychology},
  volume={66},
  number={5},
  pages={688},
  year={1974},
  publisher={American Psychological Association}
}

@article{rosenbaum1983central,
  title={The central role of the propensity score in observational studies for causal effects},
  author={Rosenbaum, Paul R and Rubin, Donald B},
  journal={Biometrika},
  volume={70},
  number={1},
  pages={41--55},
  year={1983},
  publisher={Oxford University Press}
}

@misc{chernozhukov2018double,
  title={Double/debiased machine learning for treatment and structural parameters},
  author={Chernozhukov, Victor and Chetverikov, Denis and Demirer, Mert and Duflo, Esther and Hansen, Christian and Newey, Whitney and Robins, James},
  year={2018},
  publisher={Oxford University Press Oxford, UK}
}

@article{van2007super,
  title={Super learner},
  author={Van der Laan, Mark J and Polley, Eric C and Hubbard, Alan E},
  year={2007},
  publisher={bepress}
}

@article{van2006targeted,
  title={Targeted maximum likelihood learning.},
  author={van der Laan, Mark J and Rubin, Daniel},
  journal={The International Journal of Biostatistics},
  year={2006},
  publisher={The Berkeley Electronic Press (bepress)}
}

@inproceedings{johansson2016learning,
  title={Learning representations for counterfactual inference},
  author={Johansson, Fredrik and Shalit, Uri and Sontag, David},
  booktitle={International conference on machine learning},
  pages={3020--3029},
  year={2016},
  organization={PMLR}
}

@article{de2011covariate,
  title={Covariate selection for the nonparametric estimation of an average treatment effect},
  author={De Luna, Xavier and Waernbaum, Ingeborg and Richardson, Thomas S},
  journal={Biometrika},
  volume={98},
  number={4},
  pages={861--875},
  year={2011},
  publisher={Oxford University Press}
}

@article{luo2017estimating,
  title={On estimating regression-based causal effects using sufficient dimension reduction},
  author={Luo, Wei and Zhu, Yeying and Ghosh, Debashis},
  journal={Biometrika},
  volume={104},
  number={1},
  pages={51--65},
  year={2017},
  publisher={Oxford University Press}
}

@article{ma2019robust,
  title={A robust and efficient approach to causal inference based on sparse sufficient dimension reduction},
  author={Ma, Shujie and Zhu, Liping and Zhang, Zhiwei and Tsai, Chih-Ling and Carroll, Raymond J},
  journal={Annals of statistics},
  volume={47},
  number={3},
  pages={1505},
  year={2019}
}

@article{brookhart2006variable,
  title={Variable selection for propensity score models},
  author={Brookhart, M Alan and Schneeweiss, Sebastian and Rothman, Kenneth J and Glynn, Robert J and Avorn, Jerry and St{\"u}rmer, Til},
  journal={American journal of epidemiology},
  volume={163},
  number={12},
  pages={1149--1156},
  year={2006},
  publisher={Oxford University Press}
}

@article{scholkopf2021toward,
  title={Toward causal representation learning},
  author={Sch{\"o}lkopf, Bernhard and Locatello, Francesco and Bauer, Stefan and Ke, Nan Rosemary and Kalchbrenner, Nal and Goyal, Anirudh and Bengio, Yoshua},
  journal={Proceedings of the IEEE},
  volume={109},
  number={5},
  pages={612--634},
  year={2021},
  publisher={IEEE}
}

@article{belloni2014inference,
  title={Inference on treatment effects after selection among high-dimensional controls},
  author={Belloni, Alexandre and Chernozhukov, Victor and Hansen, Christian},
  journal={Review of Economic Studies},
  volume={81},
  number={2},
  pages={608--650},
  year={2014},
  publisher={Oxford University Press}
}

@article{shortreed2017outcome,
  title={Outcome-adaptive lasso: variable selection for causal inference},
  author={Shortreed, Susan M and Ertefaie, Ashkan},
  journal={Biometrics},
  volume={73},
  number={4},
  pages={1111--1122},
  year={2017},
  publisher={Oxford University Press}
}

@article{imbens2000role,
  title={The role of the propensity score in estimating dose-response functions},
  author={Imbens, Guido W},
  journal={Biometrika},
  volume={87},
  number={3},
  pages={706--710},
  year={2000},
  publisher={Oxford University Press}
}

@article{flores2012estimating,
  title={Estimating the effects of length of exposure to instruction in a training program: The case of job corps},
  author={Flores, Carlos A and Flores-Lagunes, Alfonso and Gonzalez, Arturo and Neumann, Todd C},
  journal={Review of Economics and Statistics},
  volume={94},
  number={1},
  pages={153--171},
  year={2012},
  publisher={The MIT Press}
}

@article{huber2020direct,
  title={Direct and indirect effects of continuous treatments based on generalized propensity score weighting},
  author={Huber, Martin and Hsu, Yu-Chin and Lee, Ying-Ying and Lettry, Layal},
  journal={Journal of Applied Econometrics},
  volume={35},
  number={7},
  pages={814--840},
  year={2020},
  publisher={Wiley Online Library}
}

@article{xu2025conditional,
  title={On Conditional Stochastic Interpolation for Generative Nonlinear Sufficient Dimension Reduction},
  author={Xu, Shuntuo and Yu, Zhou and Huang, Jian},
  journal={arXiv preprint arXiv:2512.18971},
  year={2025}
}

@misc{smucler2019unifying,
      title={A unifying approach for doubly-robust $\ell_1$ regularized estimation of causal contrasts}, 
      author={Ezequiel Smucler and Andrea Rotnitzky and James M. Robins},
      year={2019},
      eprint={1904.03737},
      archivePrefix={arXiv},
      primaryClass={math.ST}
}

@book{wand1994kernel,
  title={Kernel Smoothing},
  author={Wand, M.P. and Jones, M.C.},
  isbn={9781040075173},
  series={Chapman \& Hall/CRC Monographs on Statistics and Applied Probability},
  year={1994},
  publisher={CRC Press}
}

@misc{choi2026sharpened,
      title={On Sharpened Convergence Rate of Generalized Sliced Inverse Regression for Nonlinear Sufficient Dimension Reduction}, 
      author={Chak Fung Choi and Yin Tang and Bing Li},
      year={2026},
      eprint={2602.16606},
      archivePrefix={arXiv}
}

@article{imai2004causal,
  title={Causal inference with general treatment regimes: Generalizing the propensity score},
  author={Imai, Kosuke and Van Dyk, David A},
  journal={Journal of the American Statistical Association},
  volume={99},
  number={467},
  pages={854--866},
  year={2004},
  publisher={Taylor \& Francis}
}

@article{schomaker2024causal,
author = {Schomaker, Michael and McIlleron, Helen and Denti, Paolo and Díaz, Iván},
title = {Causal Inference for Continuous Multiple Time Point Interventions},
journal = {Statistics in Medicine},
volume = {43},
number = {28},
pages = {5380-5400},
doi = {https://doi.org/10.1002/sim.10246},
year = {2024}
}

@misc{schindl2026incremental,
      title={Incremental effects for continuous exposures}, 
      author={Kyle Schindl and Shuying Shen and Edward H. Kennedy},
      year={2026},
      eprint={2409.11967},
      archivePrefix={arXiv},
      primaryClass={stat.ME}
}

@article{shin2025treatment,
author = {Heejun Shin and Antonio Linero and Michelle Audirac and Kezia Irene and Danielle Braun and Joseph Antonelli},
title = {{Treatment effect heterogeneity and importance measures for multivariate continuous treatments}},
volume = {19},
journal = {The Annals of Applied Statistics},
number = {3},
publisher = {Institute of Mathematical Statistics},
pages = {1847 -- 1867},
year = {2025},
doi = {10.1214/25-AOAS2060}
}

@Manual{Xia2025mave,
  title = {MAVE: Methods for Dimension Reduction},
  author = {Yingcun Xia and Weiqiang Hang},
  year = {2025},
  note = {R package version 1.3.12},
  url = {https://CRAN.R-project.org/package=MAVE},
  doi = {10.32614/CRAN.package.MAVE}
}

@inbook{hirano2004propensity,
author = {Hirano, Keisuke and Imbens, Guido W.},
publisher = {John Wiley \& Sons, Ltd},
isbn = {9780470090459},
title = {The Propensity Score with Continuous Treatments},
booktitle = {Applied Bayesian Modeling and Causal Inference from Incomplete‐Data Perspectives},
chapter = {7},
pages = {73-84},
doi = {https://doi.org/10.1002/0470090456.ch7},
year = {2004}
}

@book{hernan2025causal,
  title={Causal Inference: What If},
  author={Hernan, M.A. and Robins, J.M.},
  isbn={9781420076165},
  lccn={2022050839},
  series={Chapman \& Hall/CRC Monographs on Statistics \& Applied Probab},
  year={2025},
  publisher={CRC Press}
}

@article{vanderweele2013definition,
author = {Tyler J. VanderWeele and Ilya Shpitser},
title = {{On the definition of a confounder}},
volume = {41},
journal = {The Annals of Statistics},
number = {1},
publisher = {Institute of Mathematical Statistics},
pages = {196 -- 220},
year = {2013},
doi = {10.1214/12-AOS1058}
}

@book{imbens2015causal, place={Cambridge}, title={Causal Inference for Statistics, Social, and Biomedical Sciences: An Introduction}, publisher={Cambridge University Press}, author={Imbens, Guido W. and Rubin, Donald B.}, year={2015}}

@book{silverman1986density,
  title     = {Density Estimation for Statistics and Data Analysis},
  author    = {Silverman, B. W.},
  year      = {1986},
  publisher = {Chapman \& Hall},
  address   = {London},
  series    = {Chapman \& Hall/CRC Monographs on Statistics and Applied Probability},
  isbn      = {9780412246203}
}

@book{fan1996local,
  title={Local Polynomial Modelling and Its Applications: Monographs on Statistics and Applied Probability 66},
  author={Fan, J. and Gijbels, I.},
  isbn={9780412983214},
  lccn={96083481},
  series={Chapman \& Hall/CRC Monographs on Statistics \& Applied Probability},
  year={1996},
  publisher={Taylor \& Francis}
}

@article{breiman2001random,
	title = {Random {Forests}},
	volume = {45},
	issn = {1573-0565},
	doi = {10.1023/A:1010933404324},
	number = {1},
	journal = {Machine Learning},
	author = {Breiman, Leo},
	month = oct,
	year = {2001},
	pages = {5--32},
}

@book{green1993nonparametric,
  title={Nonparametric Regression and Generalized Linear Models: A roughness penalty approach},
  author={Green, P.J. and Silverman, B.W.},
  isbn={9781040072783},
  series={Chapman \& Hall/CRC Monographs on Statistics and Applied Probability},
  year={1993},
  publisher={CRC Press}
}

@article{liu2024assumption,
title = {Assumption-lean falsification tests of rate double-robustness of double-machine-learning estimators},
journal = {Journal of Econometrics},
volume = {240},
number = {2},
pages = {105500},
year = {2024},
issn = {0304-4076},
doi = {https://doi.org/10.1016/j.jeconom.2023.105500},
author = {Lin Liu and Rajarshi Mukherjee and James M. Robins}
}

@article{Botev2010Kernel,
  author = {Botev, Z. I. and Grotowski, J. F. and Kroese, D. P.},
  title = {Kernel density estimation via diffusion},
  journal = {The Annals of Statistics},
  volume = {38},
  number = {5},
  pages = {2916--2957},
  year = {2010},
  publisher = {Institute of Mathematical Statistics},
  doi = {10.1214/10-AOS799}
}

@article{marron1994visual,
  title={Visual understanding of higher-order kernels},
  author={Marron, JS},
  journal={Journal of Computational and Graphical Statistics},
  volume={3},
  number={4},
  pages={447--458},
  year={1994},
  publisher={Taylor \& Francis}
}

@article{kennedy2024semiparametric,
  title={Semiparametric doubly robust targeted double machine learning: a review},
  author={Kennedy, Edward H},
  journal={Handbook of statistical methods for precision medicine},
  pages={207--236},
  year={2024},
  publisher={Chapman and Hall/CRC}
}

@article{farrell2021deep,
author = {Farrell, Max H. and Liang, Tengyuan and Misra, Sanjog},
title = {Deep Neural Networks for Estimation and Inference},
journal = {Econometrica},
volume = {89},
number = {1},
pages = {181-213},
doi = {https://doi.org/10.3982/ECTA16901},
year = {2021}
}

@article{hornik1989multilayer,
title = {Multilayer feedforward networks are universal approximators},
journal = {Neural Networks},
volume = {2},
number = {5},
pages = {359-366},
year = {1989},
issn = {0893-6080},
doi = {https://doi.org/10.1016/0893-6080(89)90020-8},
author = {Kurt Hornik and Maxwell Stinchcombe and Halbert White}
}

@article{rotnitzky2020efficient,
  author  = {Andrea Rotnitzky and Ezequiel Smucler},
  title   = {Efficient Adjustment Sets for Population Average Causal Treatment Effect Estimation in Graphical Models},
  journal = {Journal of Machine Learning Research},
  year    = {2020},
  volume  = {21},
  number  = {188},
  pages   = {1--86},
  url     = {http://jmlr.org/papers/v21/19-1026.html}
}
\end{document}